\documentclass[preprint,12pt]{elsarticle}

\usepackage{amssymb}
\usepackage{amsmath}
\usepackage{graphicx}
\usepackage{longtable}
\usepackage{subfig}

\journal{Expert Systems with Applications}

\begin{document}

\begin{frontmatter}



\title{Unpaired Image-to-Image Translation with Content Preserving Perspective: A Review}


\author[address1]{Mehran Safayani \corref{ca1}}
\ead{safayani@iut.ac.ir}
\author[address1]{Behnaz Mirzapour}
\ead{b.mirzapour92@ec.iut.ac.ir}
\author[address1]{Hanieh aghaebrahimiyan}
\ead{hanieh.aghaebrahimiyan@ec.iut.ac.ir}
\author[address1]{Nasrin Salehi}
\ead{nasrin.salehi@ec.iut.ac.ir}
\author[address1]{Hamid Ravaee}
\ead{h.ravaee@ec.iut.ac.ir}
\cortext[ca1]{Corresponding author}
\address[address1]{Department of Electrical and Computer Engineering, Isfahan University of Technology, Isfahan 84156-83111, Iran}




\begin{abstract}
Image-to-image translation (I2I) transforms an image from a source domain to a target domain while preserving source content. Most computer vision applications are in the field of image-to-image translation, such as style transfer, image segmentation, and photo enhancement. The degree of preservation of the content of the source images in the translation process can be different according to the problem and the intended application. From this point of view, in this paper, we divide the different tasks in the field of image-to-image translation into three categories: Fully Content preserving, Partially Content preserving, and Non-Content preserving. We present different tasks, datasets, methods, results of methods for these three categories in this paper. We make a categorization for I2I methods based on the architecture of different models and study each category separately. In addition, we introduce well-known evaluation criteria in the I2I translation field. Specifically, nearly 70 different I2I models were analyzed, and more than 10 quantitative evaluation metrics and 30 distinct tasks and datasets relevant to the I2I translation problem were both introduced and assessed. Translating from simulation to real images could be well viewed as an application of fully content preserving or partially content preserving unsupervised image-to-image translation methods. So, we provide a benchmark for Sim-to-Real translation, which can be used to evaluate different methods. In general, we conclude that because of the different extent of the obligation to preserving content in various applications, it is better to consider this issue in choosing a suitable I2I model for a specific application.
\end{abstract} 









\end{frontmatter}



\section{Introduction}

\begin{figure}[tbp]
	\centering
	\includegraphics[width=1\textwidth]{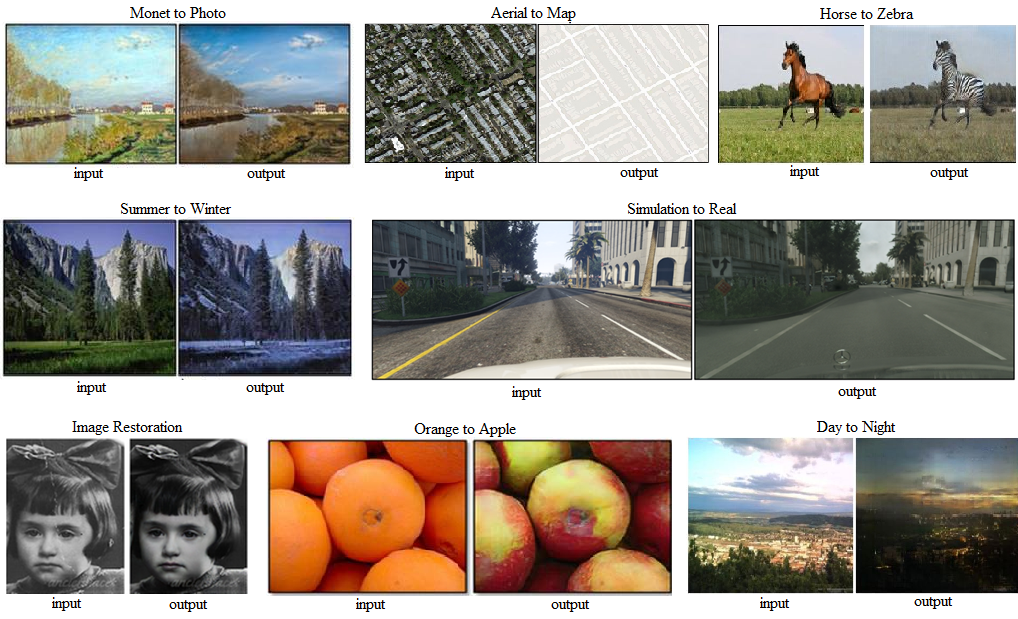}
	\caption{ 
	Several image-to-image translation problems.}
	\label{i2i_samples}
\end{figure}

Image-to-Image (I2I) translation \cite{zhu2017unpaired, isola2017image} is a task in computer vision and machine learning that involves transforming an input image from one domain (source domain) to another domain (target domain) while preserving source content and transferring target style. Some applications of I2I are artistic style transfer \cite{zhu2017unpaired, tomei2019art2real, chang2020domain}, image restoration \cite{kumar2023unpaired, cho2019dehazegan}, semantic segmentation \cite{guo2020gan, li2020simplified}, domain adaptation \cite{murez2018image, li2018heterogeneous, wilson2020survey}, virtual reality applications \cite{guo2020gan}, and so on. Converting a photo into a painting \cite{tomei2019art2real, lee2018diverse}, turning day-time images into night-time scenes \cite{yi2017dualgan}, or transforming sketches into realistic images \cite{zhu2017unpaired, huang2018multimodal}, are some examples of Image-to-Image translation tasks, as shown in Fig. \ref{i2i_samples}.

Supervised and unsupervised image-to-image translation are two different approaches of transforming images from one domain to another. Unsupervised methods do not require paired images from the source and target domains, and they learn from unaligned datasets. The lack of paired data can make it challenging to ensure high-quality translations and maintain semantic coherence in the output images. Obtaining paired datasets can be impractical for many applications, as it requires manual annotation of images \cite{pang2021image}. Hence, working image-to-image translation techniques with unpaired datasets is more common. In this paper, we focus on unsupervised I2I methods.

\begin{figure}[tbp]
	\centering
	\includegraphics[width=0.9\textwidth]{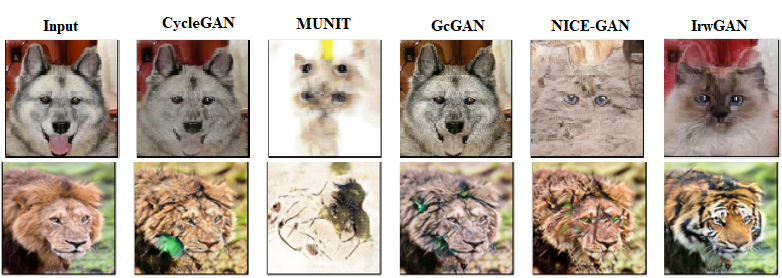}
	\caption{Visual comparisons of results of different models (CycleGAN \cite{zhu2017unpaired}, MUNIT \cite{huang2018multimodal}, GcGAN \cite{fu2019geometry}, NICE-GAN \cite{chen2020reusing}, and IrwGAN \cite{xie2021unaligned}) in preserving image content \cite{xie2021unaligned}. Tasks in order of rows: Lion2Tiger, and Dog2Cat.}
	\label{content_example_fig}
\end{figure}

Content preservation is a critical aspect of image-to-image translation \cite{chang2020domain, GDWCT, theiss2022unpaired, fan2022styleflow}, as it ensures that the essential features and structures of the original image such as objects, shapes, and relationships in the image remain intact during the transformation process. If the content is altered significantly, the resulting image may confuse viewers or not convey the intended message, leading to a lack of trust in the model's output. Content preservation is essential in various practical applications, such as medical imaging, autonomous driving, style transfer, image super-resolution, image denoising, and industrial applications. Examples to demonstrate the degree of image content preservation by different methods are shown in Fig. \ref{content_example_fig}. As it can be seen from Fig. \ref{content_example_fig}, some methods, such as MUNIT \cite{huang2018multimodal}, do not have enough power to preserve image content and have destroyed the semantics of input image. In this paper, we organize the tasks and datasets in the field of I2I translation, taking into account the content preservation perspective.

\begin{figure}[tbp]
	\centering
	\includegraphics[width=0.8\textwidth]{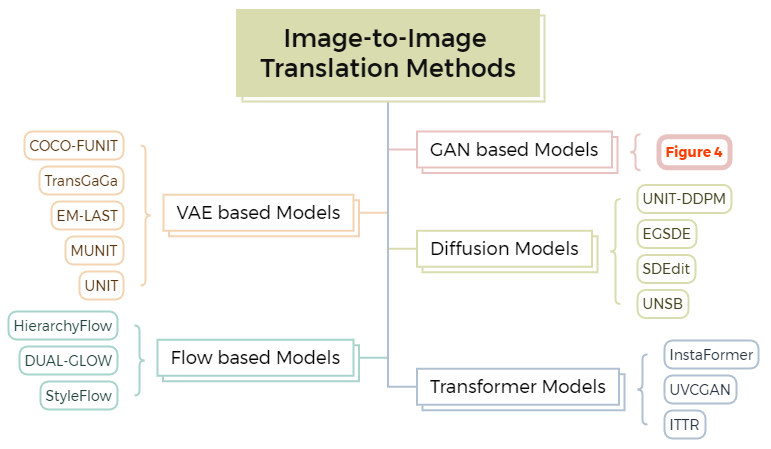}
	\caption{
	An overview of image-to-image translation methods.
	}
	\label{main_cats_papers} 
\end{figure}

In this paper, we aim to provide a comprehensive review of the recent achievements in image-to-image translation research. We categorized I2I methods into several groups based on the structure of the methods as shown in Fig. \ref{main_cats_papers}, and are discussed in the Taxonomy. Briefly, we divided previous work into 5 categories: GAN-based Models, VAE-based Models, Diffusion Models, Flow-based Models, and Transformer Models. Definitions and details of each of these structures are discussed in section \ref{Different_Models_sec}. 

\begin{figure}[tbp]
	\centering
	\includegraphics[width=0.8\textwidth]{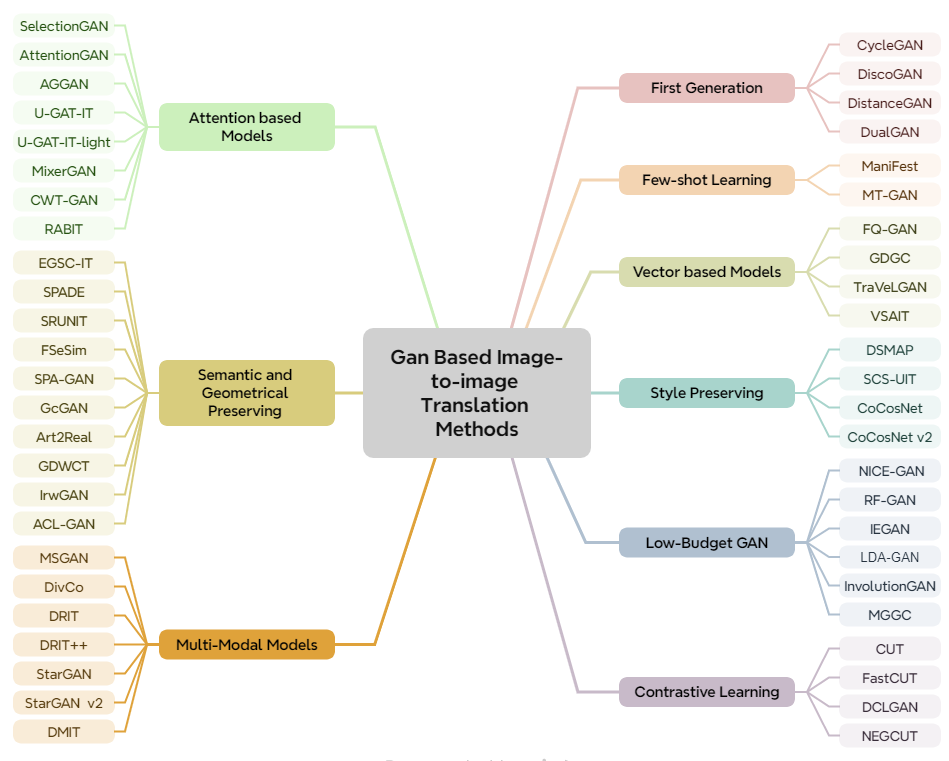}
	\caption{ 
	GAN-based image-to-image translation methods
	}
	\label{all_papers}
\end{figure}

As it can be seen from Fig. \ref{all_papers}, a large part of the previous works falls into the first category, i.e., GAN-based Models. Therefore, we have divided this category, that will be discussed in \ref{Gan_models_taxonomy}, into separate subsections. In this categorization, we have considered the innovations of works and categorized models accordingly. The sub-categories of GAN-based Models are shown in Fig. \ref{all_papers}. In addition, we have covered all the following components of the I2I translation field in this paper: different I2I tasks, datasets, evaluation metrics, performance of I2I models on different datasets and tasks based on various evaluation criteria. In total, around seventy unique I2I models were investigated. Furthermore, more than ten quantitative evaluation metrics, as well as more than thirty different tasks and datasets associated with the I2I translation challenge, were reviewed.

Image-to-image translation techniques play a crucial role in bridging the gap between simulated images generated in virtual environments and their corresponding real-world counterparts. By applying I2I models, it is possible to enhance the realism of synthetic images, making them more suitable for tasks such as training autonomous vehicles and robotic systems. In addition, the translation of simulated images to real images serves as a powerful data augmentation strategy, allowing researchers to leverage synthetic data to improve the performance of machine learning models in scenarios where real data is scarce or difficult to obtain \cite{tobin2017domain}. In this paper, we will explain datasets available and those created for simulation to real problem along with some evaluations work done on selected I2I models.

This paper is organized as follows. Section 2 introduces the popular generative models used for I2I methods. Section 3 discusses the related works on image-to-image translation. The important tasks and datasets in the I2I translation are presented in section 4. In section 5, we introduce the usual evaluation criteria for evaluating I2I translation methods. The performance of different I2I translation models on different tasks and datasets is presented in section 6. We have considered transferring simulation to real images as an application for content preserving unsupervised I2I translation benchmark, and we explain that in section 7.

\section{Different Models} \label{Different_Models_sec}
In the realm of unpaired image-to-image translation, various generative models have emerged, each with unique methodologies and underlying principles. This section delves into the prominent techniques employed in this field, including Generative Adversarial Networks (GANs), Variational Autoencoders (VAEs), Flow-based Generative Models, Diffusion Models, and Transformer Models. Each of these approaches offers distinct advantages and challenges, particularly in preserving content during the translation process. By examining these methods, we aim to provide a comprehensive understanding of their contributions to the advancement of unpaired image-to-image translation.

\subsection{Generative Adversarial Networks}
Generative Adversarial Networks (GANs) \cite{goodfellow2014generative} have emerged as a powerful framework for generating synthetic data that closely resembles real data. The core idea behind GANs is the use of two neural networks, known as the generator and the discriminator, that compete against each other in a zero-sum game. The generator's goal is to create realistic data samples, while the discriminator's task is to distinguish between real and generated samples. This adversarial process leads to the generator improving its output over time, resulting in high-quality synthetic data. Recent advancement to GANs have led to significant improvements in image synthesis \cite{zhang2018stackgan++}, text generation\cite{de2021survey, liu2020catgan}, and even video creation \cite{aldausari2022video, chen2020scripted}.
Generative Adversarial Networks encompass various models that extend the basic GAN framework. Different models of GANs have been developed to address specific challenges and enhance performance. For instance, Conditional GANs (cGANs) allow the generation of data conditioned on specific labels or inputs, enabling more controlled and targeted output. This is particularly useful in applications such as image-to-image translation, where the goal is to convert images from one domain to another while preserving certain characteristics.

\subsubsection{Unconditional GANs}
Unconditional GANs \cite{goodfellow2014generative} consist of two neural networks, a generator \( G \) and a discriminator \( D \), which are trained simultaneously through a game-theoretic approach. The generator aims to create realistic data samples from random noise, while the discriminator's role is to differentiate between real data samples and those produced by the generator. The objective function for this adversarial game can be expressed as follows:

\begin{equation}
\min_G \max_D V(D, G) = \mathbb{E}_{x \sim p_{data}(x)}[\log D(x)] + \mathbb{E}_{z \sim p_z(z)}[\log(1 - D(G(z)))]
\end{equation}

Here, \( p_{data}(x) \) represents the distribution of real data, and \( p_z(z) \) is the distribution of the noise input to the generator. The training process continues until the generator produces samples that are indistinguishable from real data, effectively leading to a Nash equilibrium \cite{goodfellow2014generative}.

\subsubsection{Conditional GANs}
Conditional GANs (cGANs) \cite{mirza2014conditional} extend the GAN framework by incorporating additional information into the generation process. This conditioning can take various forms, such as class labels, images, or other modalities. The generator \( G \) is modified to accept both the random noise \( z \) and the conditioning variable \( y \), allowing it to generate samples that are specifically tailored to the given condition. The objective function for cGANs can be represented as follows:

\begin{equation}
\min_G \max_D V(D, G) = \mathbb{E}_{x \sim p_{data}(x)}[\log D(x | y)] + \mathbb{E}_{z \sim p_z(z)}[\log(1 - D(G(z | y)))]
\end{equation}

 \begin{figure}[tbp]
	\centering
	\includegraphics[width=0.9\textwidth]{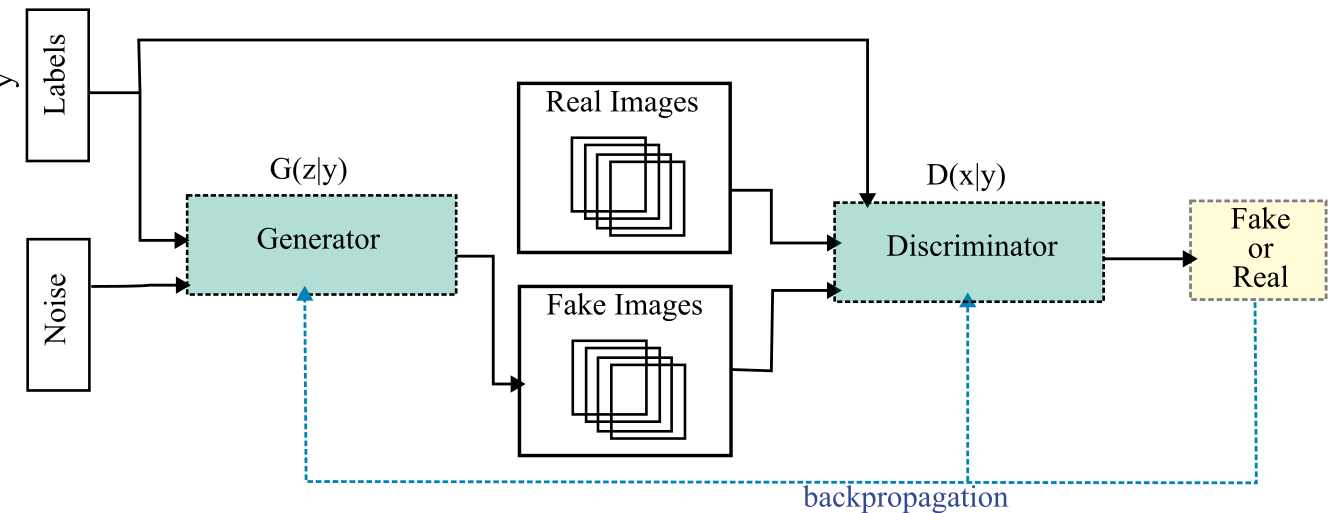}
	\caption{ 
	Conditional Generative Adversarial Network
	}
	\label{cGAN}
\end{figure}

In this formulation, the discriminator also takes the conditioning variable \( y \) into account when evaluating the authenticity of samples. This approach enables cGANs to produce more meaningful and contextually relevant outputs, making them highly effective for image generation \cite{isola2017image, gauthier2014conditional}.

\subsection{Variational AutoEncoder}
Variational Autoencoders (VAEs) \cite{kingma2013auto} combine principles from variational inference and neural networks, allowing for efficient and scalable generative modeling. The foundational concept behind VAEs is to learn a probabilistic mapping from a latent space to the data space, enabling the generation of new samples by sampling from this latent space.

The VAE framework consists of two main components: the encoder and the decoder. The encoder maps the input data \( x \) to a latent variable \( z \) by approximating the posterior distribution \( q(z|x) \). The decoder then reconstructs the input data from the latent variable by modeling the likelihood \( p(x|z) \). 

The objective of training a VAE is to maximize the Evidence Lower Bound (ELBO), which can be expressed as:

\begin{equation}
\mathcal{L}(x) = \mathbb{E}_{q(z|x)}[\log p(x|z)] - D_{KL}(q(z|x) || p(z))
\end{equation}

Here, \( \mathbb{E}_{q(z|x)}[\log p(x|z)] \) represents the expected log-likelihood of the data given the latent variables, and \( D_{KL}(q(z|x) || p(z)) \) is the Kullback-Leibler divergence between the approximate posterior \( q(z|x) \) and the prior \( p(z) \). The first term encourages the model to reconstruct the input data accurately, while the second term regularizes the latent space to follow a prior distribution, typically a standard normal distribution \( \mathcal{N}(0, I) \).

\subsection{Flow-based Generative Models}

Flow-based generative models approximate \(x \sim p^*(x)\) using samples \(\{x^{(i)}\}_{i=1}^N\). By learning an reversible mapping \(g_\theta(\cdot)\) that transforms a latent space \(p_\vartheta(z)\) into \( x \). They optimize the negative log\-likelihood, enabling direct probability density estimation \cite{pumarola2020c}.

\begin{equation}
z \sim p_\vartheta(z), \quad x = g_\theta(z),
\end{equation}

	
Normalizing flows transform a simple distribution, such as \( \mathcal{N}(z; 0, I) \), into a complex one using a sequence of invertible transformations. The bijective function \(g_\theta\) maps \(z\) to \(x\) and allows inversion to compute \(z\) from \(x\) \cite{rezende2015variational, pumarola2020c}.

\begin{equation}
	z = g_\theta^{-1}(x),
\end{equation}
	where \( g_\theta^{-1} \) is defined as a series of \( K \) invertible transformations, such that:
	\begin{equation}
	x \overset{\triangle}{=}h_0\overset{g_1^{-1}}{\leftrightarrow} h_1 \overset{g_2^{-1}}{\leftrightarrow} h_2 \cdots \overset{g_k^{-1}}{\leftrightarrow} h_K\overset{\triangle}{=} z.
	\end{equation}

	The objective of these generative models is to identify parameters \( \theta \) in such a way that the model distribution \( p_\theta(x) \) closely approximates the actual distribution \( p^*(x) \). While directly modeling \( p_\theta(x) \) is generally unfeasible, we can utilize the change-of-variables theorem to accurately calculate the log-likelihood of \( x \) within the context of the flow-based transformation \cite{pumarola2020c}:

\begin{equation}
	\log p_\theta(x) = \log p_\vartheta(z) + \log \left| \det \frac{\partial z}{\partial x} \right|,
	\end{equation}
	which can be rewritten as:
	\begin{equation}
	\log p_\theta(x) = \log p_\vartheta(z) + \sum_{i=1}^K \log \left| \det \frac{\partial h_i}{\partial h_{i-1}} \right|,
	\end{equation}
	
	The term \( \frac{\partial h_i}{\partial h_{i-1}} \) represents the Jacobian matrix of \(g_i^{-1}\), quantifying the log-density change during the transformation \cite{pumarola2020c}.

\subsection{Diffusion Models}

Diffusion models \cite{ho2020denoising} represent a category of generative models that utilize the principles of stochastic processes to create new data. They excel in producing high-dimensional outputs, including images and audio. At their core, diffusion models operate by progressively adding Gaussian noise to the training data, effectively corrupting it. The models then learn to reconstruct the original data by reversing this noising procedure. Once training is complete, data generation can be achieved by feeding randomly sampled noise into the learned denoising process. The diffusion process consists of two main phases: \\


1. Forward Diffusion Process: The forward diffusion process gradually adds noise to a data sample \( x_0 \) over \( T \) time steps. This process can be described as follows:

- At each time step \( t \) (where \( t = 1, 2, \ldots, T \)), the data is transformed according to:

\begin{equation}
x_t = \sqrt{\alpha_t} x_{t-1} + \sqrt{1 - \alpha_t} \epsilon_t
\end{equation}

where: \( \epsilon_t \sim \mathcal{N}(0, I) \) is Gaussian noise, and \( \alpha_t \) is a variance schedule that controls the amount of noise added at each step.




2. Reverse Diffusion Process: The reverse diffusion process aims to recover the original data from the noisy samples. This is modeled as a Markov chain that iteratively denoises the data:

- The reverse transition from \( x_t \) to \( x_{t-1} \) is defined as:

\begin{equation}
p_\theta(x_{t-1} | x_t) = \mathcal{N}(x_{t-1}; \mu_\theta(x_t, t), \Sigma_\theta(x_t, t))
\end{equation}

where: \( \mu_\theta(x_t, t) \) is the mean predicted by the neural network, and \( \Sigma_\theta(x_t, t) \) is the covariance, which can be set to a fixed value or learned.

3. Loss Function: The training of the diffusion model involves minimizing a loss function that measures the discrepancy between the predicted and actual data distributions. A common choice is the variational lower bound, which can be expressed as:

\begin{equation}
L(\theta) = \mathbb{E}_{t, x_0, \epsilon} \left[ \left\| \epsilon - \epsilon_\theta(x_t, t) \right\|^2 \right]    
\end{equation}

where: \( \epsilon_\theta(x_t, t) \) is the model's prediction of the noise at time \( t \), and The expectation is taken over the time step \( t \), the original data \( x_0 \), and the noise \( \epsilon \).

Integrating diffusion models into image-to-image translation frameworks presents significant enhancement in the quality, diversity, and adaptability of the models \cite{sasaki2021unit, zhao2022egsde, meng2022sdedit, kim2023unpaired}. The iterative nature of diffusion models allows for smoother transitions between the input and the output images. 
Furthermore, by applying stochastic processes during the denoising steps, diffusion models can produce diverse outputs for the same input. Also, diffusion models can learn to represent multiple modes in data, which is useful in many applications where the output could have multiple versions such as in the case of multi-modal outputs.

\subsection{Transformers Models}
Transformers~\cite{vaswani2017attention}, originally developed for natural language processing (NLP) tasks, have recently been adapted to various computer vision applications, including image-to-image translation. Unlike traditional convolutional neural networks (CNNs), transformers are particularly adept at capturing non-local patterns within images due to their efficient attention mechanisms. This capability has led to the emergence of Vision Transformers (ViTs)~\cite{alexey2020image} and hybrid CNN-transformer models~\cite{ranftl2021vision}, which have demonstrated superior performance over previous models in several computer vision tasks.

Transformers are a type of neural network architecture composed of multiple layers. Each layer includes a multi-head self-attention mechanism, parallel fully-connected networks, residual connections, and layer normalization. When given a sequence of $N$ input embeddings, transformers generate a sequence of $N$ output embeddings, where each output embedding captures the relationship between its corresponding input embedding and the entire input sequence~\cite{han2022survey}. The self-attention mechanism enables each input embedding (each row of $X$) to simultaneously consider all other embeddings in the input sequence. It calculates an attention score for each pair of inputs by projecting each input into a query tensor $Q = XW_q$ and a key tensor $K = XW_k$. The attention scores are obtained by taking the dot product of each query vector (a row of the query tensor) with every key vector (a row of the key tensor), followed by a softmax operation that normalizes these scores so that they sum to one for each query. These attention scores are then used to compute a weighted sum of value tensors $V = XW_v$.
\begin{equation}
Attention(Q,K,V) = softmax (Q.K^T/\sqrt{d_q})\times V
\end{equation}

To stabilize gradients during training, the dot product is scaled by a factor of $\sqrt{d_q}$, where $d_q$ is the dimension of the query tensor. Transformers typically compute multiple sets of attention scores in parallel, each with its own set of learned parameters $W_q$, $W_k$, and $W_v$. This allows the model to focus on different aspects of the input sequence simultaneously, a process known as multi-head self-attention. The outputs of each attention "head" are concatenated and linearly transformed to create the final output representation. In tasks where the order of inputs is crucial, a positional encoding is added to help the network recognize the position of each input in the sequence~\cite{khan2022transformers}.

Vision Transformers (ViTs)~\cite{dosovitskiy2020image}, initially designed for recognition tasks such as image classification, object detection, and semantic segmentation, have also been explored for image generation. Despite their success in recognition based applications, directly applying ViTs to image-to-image translation presents challenges due to the complexity of generating high-quality images while maintaining semantic consistency. To overcome these challenges, transformer-based generator architectures have been proposed, capable of capturing long-range dependencies through multi-head self-attention blocks. However, these adaptations often face significant computational and memory demands, necessitating further optimization for image generation tasks.


\section{Taxonomy}

The purpose of this paper is on unpaired image to image translation methods. These methods aim at learning a mapping of images from a source domain to a target domain while preserving the content of the input image, while the model does not require paired images.

In this section, the main categorization of the studied I2I translation methods is based on the architectural structure of the models, and hence they are placed in 5 categories: GAN-based Models, VAE-based Models, Diffusion Models, Flow-based Models, and Transformer Models. In summary, GANs \cite{goodfellow2014generative} consist of two main components—a generator and a discriminator—which compete against each other. The generator creates images, while the discriminator evaluates their realism. VAEs \cite{kingma2013auto} consist of an encoder that maps input data to a latent space and a decoder that reconstructs images from this latent representation. Diffusion models \cite{ho2020denoising} are probabilistic models that convert data into noise through a forward diffusion process and then learn to reverse this process to generate new data. Flow-based models \cite{weng2018flowmodels} utilize a series of invertible transformations that map data from a simple distribution to a complex distribution. This is achieved through a sequence of "flows" that apply learned transformations while ensuring the overall transformation remains invertible. Transformer models \cite{vaswani2017attention} adopts self-attention modules to capture global context information, thus improving instance level awareness while translating out content features regarded as tokens.

Since most of the existing works are placed in the first category, this category has a larger volume and is itself divided into subsections. First Generation, studies pioneer and basic I2I models. Semantic and Geometrical Preserving, refers to the degree to which the generated images maintain the underlying semantic information present in the input images, and the property that simple geometric transformations do not alter the semantic structure of images, respectively. Style Preserving, involves taking the visual appearance or style of one image and applying it to the content of another image, while preserving the essential structure and content of the original image. The central idea of Contrastive Learning is the formation of positive pairs and negative pairs on images. Then, the model learns important features that help in translation by maximizing the similarity of the positive pairs while minimizing that of the negative pair.
Few-shot Learning, is used to perform effective translations when only a few reference images from the target domain are available. Low-Budget GAN, refers to models that are to resolve the problem of high computational cost and complexity of GAN-based image translation methods. Vector based Models, refers to methods that transfer the features of input images to an intermediate vector space and continue calculations with them. Attention based Models, are deep learning techniques used to allow the model to focus on specific parts of an image when generating outputs. Multi-Modal Models, refer to frameworks that can generate multiple possible outputs from a single input image.

\subsection{GAN based Models} \label{Gan_models_taxonomy}

\subsubsection{First Generation}
The pioneering first-generation GAN-based models laid the foundation for unpaired image-to-image translation, tackling the challenges of learning mappings between domains without paired data while preserving semantic and structural coherence.

CycleGAN~\cite{zhu2017unpaired} is a pioneering model for unpaired image-to-image translation, learning mappings between two domains $X$ and $Y$ without requiring paired images. It introduces two mappings, $G:X \to Y$ and $F:Y \to X$, along with adversarial discriminators $D_X$ and $D_Y$, which distinguish real from generated images. The model optimizes adversarial losses to make generated images resemble the target domain and enforces cycle consistency to ensure an image translated to the other domain and back remains similar to the original. Additionally, identity loss regularizes the generator to maintain essential features when translating images within the same domain. CycleGAN has been widely used in applications like photo-to-painting conversion and style transfer, showcasing its effectiveness in unpaired translation tasks.

The objective of CycleGAN, as shown in Equation \ref{eq_cyclegan}, combines adversarial losses for both generators, $\mathcal{L}_{GAN}(G, D_Y, X, Y)$ and $\mathcal{L}_{GAN}(F, D_X, Y, X)$, which encourage the generators to produce images indistinguishable from the target domain. Additionally, the cycle consistency loss, $\mathcal{L}_{cyc}(G, F)$, ensures that an image mapped from one domain to another can be translated back to the original domain. The identity loss, $\mathcal{L}_{identity}(G, F)$, regularizes the generators to preserve features when translating images from the target domain. Hyperparameters $\lambda$ and $\beta$ control the balance of the cycle consistency and identity losses, respectively.

\begingroup\makeatletter\def\f@size{9}\check@mathfonts
\begin{equation}
\mathcal{L}(G, F, D_X, D_Y) = \mathcal{L}_{GAN}(G, D_Y, X, Y) 
+ \mathcal{L}_{GAN}(F, D_X, Y, X) 
+ \lambda \mathcal{L}_{cyc}(G, F)
+ \beta \mathcal{L}_{identity}(G, F),
\label{eq_cyclegan}
\end{equation}
\endgroup



Another pioneer in unpaired image-to-image translation is DiscoGAN~\cite{kim2017learning}, a model designed to learn relationships between two domains without paired examples by training two generators for bidirectional mapping. It enforces a one-to-one correspondence through a reconstruction loss, ensuring that translating an image to another domain and back preserves its original structure. The architecture consists of two coupled GANs with shared parameters and separate discriminators, balancing adversarial and reconstruction losses to generate realistic images while maintaining invertibility. This approach prevents mode collapse and ensures meaningful cross-domain transformations.

The main difference between DiscoGAN and CycleGAN lies in the training objective and emphasis on discovery. While both models employ bidirectional generators and use reconstruction losses to enforce a cycle consistency, DiscoGAN specifically focuses on discovering one-to-one correspondences between unpaired data, making it particularly useful for relation discovery between two domains. In contrast, CycleGAN is more general and designed for various tasks where paired data is unavailable, such as image translation. DiscoGAN’s stricter bijection constraint distinguishes it in tasks requiring exact mapping across domains.

DistanceGAN~\cite{benaim2017one} is a model for unsupervised image-to-image translation that focuses on one-sided mapping from domain $A$ to domain $B$ without requiring paired datasets. Unlike bidirectional approaches, it enforces a distance-preserving constraint, ensuring that visually similar images in domain $A$ remain similarly spaced in domain $B$. This helps maintain consistency and realism while reducing mode collapse. Additionally, DistanceGAN employs adversarial loss to align generated images with the target distribution and introduces a "self-distance" constraint to preserve internal structural relationships. This approach enables effective translation between structurally dissimilar objects, such as transforming shoes into edge maps or mapping handbags to shoes.

DualGAN~\cite{yi2017dualgan} belongs to first-generation GAN-based approaches to unsupervised image-to-image translation, enabling the mapping between two unpaired and unlabeled domains, U and V, through dual learning. It trains two GANs simultaneously, one for translating images from U to V and another for the reverse direction, ensuring consistency by enforcing a cycle of translations that minimizes reconstruction errors. To improve stability and avoid mode collapse, DualGAN employs Wasserstein GAN (WGAN)~\cite{arjovsky2017wasserstein} loss and incorporates L1 distance to maintain domain distribution while preventing image blurriness. Its architecture includes U-Net-like skip connections to preserve object details and PatchGAN~\cite{isola2017image} discriminators, which focus on local textures and high-frequency details, improving the realism of generated images while keeping the model computationally efficient.

\subsubsection{Semantic and Geometrical Preserving}
EGSC-IT \cite{ma2018exemplar} presents a novel approach to achieving semantic consistency in unsupervised image-to-image translation through the Exemplar Guided \& Semantically Consistent Image-to-image Translation (EGSC-IT) framework. Traditional methods often rely on semantic labels, which are difficult to obtain, leading to inconsistencies during translation. Instead, this model utilizes feature masks as attention modules to approximate semantic categories without requiring labeled data, thus maintaining semantic integrity during the translation process.The framework employs Adaptive Instance Normalization (AdaIN) to apply style from target domain exemplars while ensuring that the shared content component remains semantically consistent. By computing feature masks, the model effectively decouples different semantic categories, preventing the mixing of styles from various objects and scenes.

The  goal of SPADE model presented in \cite{park2019semantic} is to generate photorealistic images from segmentation masks. Traditional normalization layers often wash away critical semantic information, leading to suboptimal results. To address this, the authors propose a novel spatially-adaptive normalization method that modulates activations based on the input semantic layout, effectively preserving semantic details throughout the network. SPADE allows for spatially adaptive transformations, ensuring that different regions of the image maintain their semantic integrity. This approach not only enhances the quality of the generated images but also supports multi-modal synthesis, enabling diverse outputs from the same input mask.

SRUNIT \cite{jia2021semantically} addresses the challenge of preserving semantic consistency during unpaired image translations. Traditional methods often suffer from semantics flipping, where the meaning of the input images is altered in the output. This model enhances semantic robustness by ensuring that perceptually similar contents are mapped to semantically similar outputs, thus preventing the flipping issue.The method employs a semantic robustness loss that minimizes the distance between the semantics of the input and translated images, focusing on multi-scale feature space perturbations. This approach contrasts with existing distance-preserving methods, which can lead to artifacts and reduced diversity in translations.


The FSeSim model \cite{zheng2021spatially} introduces a novel spatially-correlative loss for image-to-image (I2I) translation, focusing on preserving scene structure while allowing for significant appearance changes. Traditional methods often struggle with maintaining content integrity due to their reliance on pixel-level or feature-level losses. This method proposes self-similarity patterns using Fixed Self-Similarity (FSeSim) and Learned Self-Similarity (LSeSim) losses to define scene structure. FSeSim uses fixed features to create spatially-correlative maps, while LSeSim employs self-supervised contrastive learning to enhance adaptability across domains. The approach is integrated into existing architectures, replacing cycle-consistency loss with the proposed losses, and demonstrates superior performance in maintaining structural fidelity across various tasks, including single-modal, multi-modal, and single-image translations.

SPA-GAN \cite{emami2020spa} is a novel spatial attention GAN model designed for image-to-image translation tasks, incorporating an attention mechanism directly into the GAN architecture. This allows the generator to focus on discriminative regions between source and target domains, resulting in more realistic images. The model computes spatial attention maps in the discriminator and transfers this knowledge to the generator, enhancing image quality. In this model, an additional feature map loss is introduced to preserve domain-specific features during translation, ensuring realistic outputs.

GcGAN \cite{fu2019geometry} presents a novel approach for unsupervised domain mapping using Geometry-Consistent Generative Adversarial Networks (GcGAN). This method focuses on one-sided domain mapping, allowing the learning of a mapping function without its inverse. The key innovation is the geometry-consistency constraint, which leverages the property that simple geometric transformations do not alter the semantic structure of images. By incorporating this constraint, GcGAN reduces the solution space while preserving correct solutions, leading to more accurate domain mappings. In the GcGAN framework, vertical Flipping and rotation were utilized as geometric transformations to demonstrate the effectiveness of the geometry-consistency constraint.

Tomei et al. in \cite{tomei2019art2real} propose Art2Real, a framework for transforming artistic images into photo-realistic visualizations while preserving the semantics of the artwork. A number of memory banks of realistic patches are built from real photos, each containing patches from a single semantic class. By comparing generated and real images at the patch level, a generator learns to generate photo-realistic images, while preserving the semantics of the original painting. This model extracts fixed-size patches from the set of real images through a sliding-window policy and puts them in a corresponding memory bank. Art2Real combines cycle-consistent adversarial loss with semantically-aware translation loss. This hybrid strategy is advantageous to the model since it allows for the creation of images that is both visually pleasing and content correct.

Cho et al. in \cite{GDWCT} propose GDWCT, an approach to I2I translation by leveraging a group-wise deep whitening-and-coloring transformation (WCT). GDWCT is a two-step transformation process that consists of whitening (which normalizes the input image features) and coloring (changing the covariance matrix of the whitened feature to those of the style feature) operations. Therefore, the authors add the GDWCT module in each residual block of the generator. By injecting the style information across multiple points of generative model via a sequence of GDWCT modules, the translation model can simultaneously reflect both the fine and coarse-level style information, while preserving semantic content and ensuring high-quality output.

Xie et al. in \cite{xie2021unaligned} propose IrwGAN, an approach to address the problem of unaligned I2I translation. They offer a reweighting mechanism which learns to modify the significance of different samples during the training process. This technique helps the model pay attentions to the source domain samples which are most helpful to the translation task. With this reweighting mechanism, the model is able to model the connections between the source and target domains more effectively. This method incorporates the reweighting process into the training process. The loss function is similar to the loss used in other I2I models, with the difference that it has an additional loss called effective sample size loss. Since there is no constraint on importance weights for aligned images, they propose the effective sample size loss to control these weights and allow more aligned images to be selected for translation. 

The ACL-GAN model \cite{zhao2020unpaired} aims to translate images between two domains. It uses two generators, $ G_S $ and $ G_T $, to translate images to the source ($ X_S $) and target ($ X_T $) domains, respectively, with each generator comprising a noise encoder, image encoder, and decoder. The generators accept pairs of an image and a noise vector, with the noise vector used for identity loss, ensuring the output image retains specific features of the input. Additionally, discriminators $ D_S $ and $ D_T $ differentiate between real and generated images, while a consistency discriminator ensures that translated images stay consistent with the source.

\subsubsection{Content Preserving} \label{Style_sec}
Image-to-image transformation models in style transfer strive to preserve the content of the original image, such as its shape and structure, while incorporating the style of the target image. These models blend the visual features of the target domain, such as colors and textures, into the input image while maintaining the original arrangement and structure of objects. The key challenge lies in striking the right balance between preserving the original content and integrating the target style, ensuring the final output not only strongly resembles the style of the target domain but also retains the structural details of the original image.


 DSMAP\cite{chang2020domain} proposes a training framework aimed at improving the quality of image style transfer by utilizing mappings tailored to specific domains. This framework addresses the limitations of previous methods that rely on a shared domain-invariant content space, which can compromise content representation. By remapping shared latent features into domain-specific content spaces, images can be encoded more effectively for style transfer. This method involves the development of domain-specific content and style encoders that work alongside an adversarial network consisting of a generator and a discriminator. This allows the generator to blend content and style features from different domains, producing more realistic and aesthetically pleasing images. A key advantage of DSMAP is its ability to handle challenging scenarios that require semantic correspondences between images, significantly improving image quality and style alignment.

The SCS-UIT model\cite{liu2021separating} includes encoders for feature extraction, generators for image creation, and detectors for quality evaluation, focusing on content separation and style codes. Its content-style separator consists of residual blocks and a DI-HV (Domain-Invariant High-Level Vision) task mapping function, which is connected to the Squeeze-Selection-and-Excitation (SSE) module. 
DI-HV, which plays a key role in this model, refers to tasks that identify common features across different domains. In this model, the content features of domain A are separated from the style features, and these features are then used for translation to domain B. These features, especially when working with data from different domains, help the model preserve the main content of the image regardless of domain differences. The SSE module learns feature channel weights to determine their importance. It selects high-scoring units as content codes and labels the other units as style codes.

In this paper, NAIN (Normalized Adaptive Instance Normalization) is proposed to improve image quality by normalizing the parameters of Adaptive Instance Normalization (AdaIN) \cite{huang2017arbitrary} and preventing the occurrence of water-drop effects. This process also controls the intermediate values, preventing quality degradation. The model uses four loss functions to achieve its objectives: one compares the input with the reconstructed images, another enhances quality through a GAN framework, a third evaluates feature similarity with reference images, and the fourth assesses performance on common visual tasks. Together, these functions help identify shared information across domains while reducing the need for labeled data.

CoCosNet is a GAN-based model for translating images between domains A and B using a style reference. It takes an input image $x_A \in A $ and a reference image $ y_B \in B $, aiming to retain the content of $ x_A $ while adopting the style from semantically similar parts of $ y_B $. The model first establishes a semantic correspondence between the two images, aligning the reference image with the source. This is achieved through a cross-domain correspondence network that maps both images into a shared domain SS, using a feature pyramid network and domain-specific transformations. A correspondence layer computes a correlation matrix to match semantically similar regions. The translation network then generates the styled output by injecting style information from the reference into the source image using spatially-adaptive denormalization (SPADE) layers. This process preserves the structural consistency of the source image while applying the style. By combining positional normalization (PN) and SPADE, the network ensures high-quality texture transfer, producing realistic and style-consistent results.

CoCosNet v2 \cite{zhou2021cocosnet} improves the original model by introducing multi-level domain alignment and refined correspondence matching for better alignment across domains. It constructs a latent space pyramid, capturing details from low to high resolution while preserving contextual information through U-Net architecture and skip connections. A GRU-assisted PatchMatch algorithm iteratively refines feature correspondences across pyramid levels, combining hierarchical alignment with coarse-to-fine matching for highly accurate correspondence fields. Additionally, a differentiable warping function aligns warped exemplars to the target domain, enhancing output quality. The translation network utilizes SPADE with adaptive modulation, ensuring structural consistency and style fidelity in the generated images.


\subsubsection{Contrastive Learning}

Contrastive Learning is a technique that enhances the performance of vision tasks by forming of positive pairs and negative pairs on data. The model will be trained to emphasize the features helpful in translation by maximizing the similarities of positive pairs yet minimizing the correlations between negative pairs.

Park et al. in \cite{CUT} propose CUT to use contrastive learning techniques to establish relationships between unpaired images. This involves learning representations that can differentiate between similar and dissimilar images. Their method is to maximize mutual information between the two domains, in a patch-based approach. CUT promotes the mapping of corresponding patches to similar points within a learned feature space, in relation to other patches in the dataset. This method focuses on maintaining the semantic content of images while allowing for flexibility in style transfer. They also proposed the FastCUT method. FastCUT builds upon the foundation laid by CUT but aims to improve computational efficiency and speed. It optimizes the contrastive learning process, potentially by reducing the number of negative samples or using more efficient sampling strategies. While both methods CUT and FastCUT are based on contrastive learning, FastCUT enhances efficiency compared to the original CUT approach.

Han et al. in \cite{han2021dual} propose the DCLGAN method, an approach that leverages dual contrastive learning to improve the quality and robustness of unsupervised I2I translation tasks. DCLGAN employs two contrastive learning objectives simultaneously, one focusing on the content of images and the other on their style. This dual approach helps in better disentangling the content and style representations, leading to improved translation quality. By utilizing this approach, the model becomes more robust to variations in input data and better preservation of semantic content across different styles.

Wang et al. in \cite{NEGCUT} propose NEGCUT, a strategy to enhance contrastive learning by focusing on the generation of hard negative examples. The main contribution of this work is the development of a method to generate hard negative examples that are specifically challenging for the model. By including these challenging examples in the training pipeline, the model becomes better at capturing the nuances of content and style in unpaired image-to-image translation tasks. In addition, unlike traditional methods, this approach generates hard negatives on an instance-by-instance basis. The incorporation of instance-wise hard negatives leads to superior performance compared to standard contrastive learning approaches.

\subsubsection{Few-shot Learning}

When only a small number of reference images from the target domain are available, applying few-shot learning approaches in the field of I2I translation can be effectively helpful.


Pizzati et al. in \cite{Manifest} propose ManiFest, an approach to few-shot image translation by employing manifold deformation techniques. This method considers that images can be represented as points on a manifold in a space. This perspective allows for a more structured understanding of how images relate to one another. The manifold deformation mechanism allows the model to adaptively manipulate the learned manifold by deforming it based on a few reference images from the target domain. This deformation helps capture the unique style and characteristics of the target images. This process is designed to ensure that the semantic content of the source images while applying the target domain's style, remains intact.

Lin et al. in \cite{lin2020learning} propose the Meta-Translation GAN (MT-GAN), an approach to unsupervised domain translation by leveraging meta-learning techniques. During the meta-training process, MT-GAN trained on both a primary translation task and a synthesized dual translation task, incorporating a cycle-consistency meta-optimization objective to enhance generalization. The effectiveness of MT-GAN demonstrated across diverse two-domain translation tasks, showing significant improvements over existing methods, especially when each domain has limited training samples.

\subsubsection{Low-Budget GAN}

In order to resolve the problem of high computational cost and complexity of GAN-based image-to-image translation methods, some efforts have been made to compress GAN-based models, reduce the computational cost, and make the models more compatible with limited hardware resources.


Chen et al. in \cite{chen2020reusing} propose NICE-GAN, an unsupervised I2I translation method that enhances encoding by reusing discriminators for both adversarial training and feature extraction. This approach capitalizes on the discriminators' ability to perform domain-specific feature extraction, enabling effective translation of relevant features from the source to the target domain while maintaining semantic consistency. By reusing discriminators, NICE-GAN achieves a more compact architecture suitable for limited budgets. Additionally, it employs multi-scale discriminators, dividing the classifier into three sub-classifiers to facilitate processing at local, middle, and global scales.

RF-GAN \cite{koksal2020rf} is a generative adversarial network designed to address the challenges of high model complexity and data requirements in image-to-image translation tasks. RF-GAN significantly reduces the number of parameters up to 75 percent less than state-of-the-art GANs by employing a single re-configurable generator and a multi-domain discriminator, which allows it to operate effectively in low-budget and resource-constrained environments. This innovative architecture enables RF-GAN to perform high-fidelity translations across two or multiple domains without the need for additional classifiers or multiple discriminators, which are common in existing models. As a result, RF-GAN can be trained with fewer images while still achieving high performance, making it particularly suitable for scenarios where data is limited. The ability to maintain high translation fidelity with a compact model size not only enhances its applicability in edge computing but also facilitates its use in various practical applications where computational resources are constrained. Also, iterative training process employed by RF-GAN ensures that the generator can learn mappings in both directions (S → T and T → S) without confusion, maintaining cycle consistency and improving overall translation quality.


Ye et al. in \cite{ye2021independent} propose IEGAN, a framework that employs independent encoders for different domains and a hierarchical structure. The common GAN models includes two encoders, one for generator and another for discriminator. Since the input of each network is an image and there are only two image domains, there is no need for more encoders. So they propose to remove the encoder of each network and establish an encoder as an independent encoder for both networks. In this way, the encoder focuses on learning the features of the input image, thereby ensuring the encoding capacility of encoder. The use of independent encoders for each domain allows the model to learn distinct feature representations that capture the unique characteristics of the respective domains and focus on maintaining semantic consistency and reducing artifacts in the generated images. Also, this model is more compact compared to normal GAN based methods, and less computational cost is expected.

Translation strategies should focus on important object regions of an image instead of non-critical areas like the background to simplify mapping learning challenges. Based on this concept, Zhao et al. in \cite{ZHAO2022355}, focus on context-aware I2I translation and propose a lightweight domain-attention generative adversarial network (LDA-GAN) model which requires significantly lower resources for translating images adequately. They propose an enhanced domain attention module (DAM) that allows the generator to direct attention to more important regions of the images being translated, leading to more believable translation outcomes. Additionally, a pretrained VGG-16 encoder is utilized to transform images into feature space, minimizing irrelevant details. The introduction of a separable-residual block (SRB) preserves spatial information and enhances detail accuracy during translation.  Both these two enhancements are able to decrease the number of the model learnable parameters.

Deng et al. in \cite{deng2023involutiongan} improve the GAN architecture by using a light-weight operator termed involution that helps in utilizing not just local features but also long-range dependency between channels, and serves as an alternative to convolution in neural networks. This approach reduces the number of parameters and computational complexity while maintaining performance. They also noted that feature-level reconstruction discrepancy between original and reconstructed images is significant, and such information is meaningful to the enhancement of the reconstructed image’s quality. Hence, they introduce an additional loss term named LPIPS loss that guides the training by assessing the learned perceptual similarity distance. InvolutionGAN achieves competitive results compared to existing state-of-the-art methods but with a significantly lighter model.

Ganjdanesh et al. in \cite{ganjdanesh2024compressing} address the challenge of reducing the computational cost of GANs used for I2I translation without sacrificing output quality. This paper introduces an approach, called MGGC (Manifold Guided GAN Compression), to GAN compression that focuses on preserving this density structure in the pruned model. The method involves partitioning the learned manifold into local neighborhoods around generated samples and proposing a pruning objective that regularizes the pruned model to maintain the local density structure. Additionally, a collaborative pruning scheme is developed, where both the discriminator and generator are pruned by agents that exchange feedback to optimize their architectures. This design enables the identification of effective sub-networks while ensuring a balanced pruning process between the generator and discriminator. The results of using MGGC along with the CycleGAN can be seen in section \ref{Evaluation_sec} as CycleGAN-MGGC.

\subsubsection{Vector based Models}

In vector-based methods, the entire vector represents the entire structure or image. This can be advantageous for capturing global patterns and relationships in images, which are critical for effective translation.

Amodio et al. in \cite{Travelgan} propose TraVeLGAN, an approach to unsupervised I2I translation that focuses on learning transformation vectors to facilitate the translation process. The core innovation of this model is the concept of transformation vectors, which represent the changes needed to translate images from one domain to another. By learning these vectors, the model can effectively capture the relationships between different image styles or contents. The authors add a siamese network \cite{chicco2021siamese} to two-network system of original GAN, guiding the generator to ensure that each original image retains semantic similarity with its generated counterpart. In addition, to ensure that the translations are coherent and preserve the original content, the model incorporates cycle consistency loss. 

Theiss et al. in \cite{theiss2022unpaired} propose VSAIT, an approach to unpaired I2I translation by leveraging vector symbolic architectures (VSA) to facilitate the transformation of images between domains. VSA allows for the representation and manipulation of high-dimensional vectors that can encode complex relationships and structures. The method utilizes VSA to encode semantic information from input images, by representing images as high-dimensional vectors. The authors propose specific loss functions that work in conjunction with the VSA framework to ensure that the translated images maintain coherence and quality. This includes losses that emphasize both style and content preservation.

Zhao et al. in \cite{FQ-GAN} propose FQ-GAN, an approach to enhance the training of GANs through feature quantization for I2I translation. Feature Quantization involves discretizing continuous feature representations into a finite set of values, which can simplify the learning process, and leading to improved performance in generating high-quality images. One of the key contributions of the paper is demonstrating that feature quantization enhances the stability of GAN training, leading to higher quality image generation. By mitigating the variance in the feature space, the model is less prone to mode collapse and can converge more reliably during training.

The paper GDGC \cite{cao2023guided} presents a guided I2I translation model that enhances communication between the generator and discriminator in GANs. By reformulating the generator's training as a Partially-observed Markov Decision Process (POMDP), the model addresses the limitations of existing frameworks, particularly in unpaired translation scenarios. The proposed guidance mechanism allows for dynamic learning of continuous communication vectors, improving the generator's performance without the constraints of traditional methods. The model integrates cycle consistency loss to maintain content integrity while transforming textures, and it employs various merging strategies for guidance vectors to optimize information flow.

\subsubsection{Multi-Modal Models}

Multi-domain image-to-image translation models are designed to take a single input and transform it into multiple different domains. These models can translate the input image into various forms or states.

StarGAN\cite{choi2018stargan} is a multi-domain image-to-image translation framework that enables training on multiple datasets within a single network. It translates images based on target domain labels and uses an auxiliary classifier to manage domains. The generator extracts domain-specific features, while the discriminator classifies domain labels. StarGAN v2\cite{choi2020stargan} enhances diversity and scalability across domains with four key modules: a generator, a mapping network for style codes, a style encoder, and a discriminator. It improves upon previous models with adversarial loss for realism, style reconstruction for consistency, style diversification for variation, and cycle consistency to preserve input characteristics.

The DRIT++ paper focuses on learning a multimodal mapping between two or more visual domains without requiring paired data. The model utilizes content and feature encoders, generators, and domain discriminators to achieve this mapping. It swaps feature representations between dissimilar images, with the content encoder mapping shared information into a common content space. This allows the model to combine the content of one image with the features of another, generating diverse outputs.To ensure accurate reconstruction, cross-cycle consistency loss is applied, while a content discriminator separates content features from domain-specific features. Additionally, mode-seeking regularization prevents convergence to a single mode, enhancing output diversity. This framework also extends to multi-domain translation, where images from multiple domains are processed using a shared encoder and a common generator for all domains.

The DMIT\cite{yu2019multi} model aims to separate content and style in images using an encoder-decoder architecture and conditional adversarial training in the feature space. It divides images into latent content and style spaces, where the content encoder extracts domain-invariant content, and the style encoder captures domain-specific style. The generator (G) uses residual blocks and deconvolutional layers to generate images, with a unified discriminator applied across all domains.The model has two main paths: the disentanglement path uses cVAE and adversarial training to separate style and content, and the translation path encourages diverse image generation by sampling domain labels and styles. The loss function combines disentanglement and translation losses to improve both image quality and diversity.

DivCo\cite{liu2021divco} focuses on generating diverse conditional images using conditional Generative Adversarial Networks (cGANs). One of the main challenges faced by cGANs is mode collapse, where changes in the latent codes do not lead to sufficient diversity in the generated images. To address this issue, DivCo employs a latent-augmented contrastive loss that takes into account both positive and negative relationships among the generated images. The generator combines a specific condition with a latent code sampled from a Gaussian distribution to create images. The objective is to ensure that images generated from nearby latent codes are visually similar, while those produced from more distant latent codes exhibit significant visual differences. This contrastive approach enables DivCo to produce a wider range of images by leveraging multiple modes, effectively mitigating the risk of mode collapse.

The Mode Seeking GANs (MSGANs)\cite{mao2019mode} method is introduced to address the issue of mode collapse in conditional Generative Adversarial Networks . The main idea behind this approach is that if different latent codes are provided to the model, the generated images should exhibit significant diversity. By adding a new regularization term to the objective function, the generator is encouraged to maximize the distance between images generated from different latent codes relative to the distance between those codes. This approach motivates the generator to explore less represented modes and increases the diversity of generated samples. The proposed method is easily applicable to various cGAN models.

\subsubsection{Attention based Models}

The attention based I2I model takes into account the attention mechanism to prioritize certain image features or regions. It assigns different weights to different parts of the feature map for the model to concentrate on the most relevant areas concerning task translation.

Tang et al. in \cite{SelectionGAN} propose SelectionGAN, a framework for cross-view image translation that enhances the quality of generated images through a multi-channel attention mechanism and cascaded semantic guidance. Translating images across different views is a challenging task. The multi-channel attention mechanism allows the network to selectively focus on different aspects of the input images. SelectionGAN uses a selection process to choose the best generated images for further processing, enhancing output quality. It also employs cascaded semantic guidance to utilize semantic information at multiple levels, resulting in high-quality translations that are visually appealing and semantically coherent.

Tang et al. also propose AGGAN in \cite{AGGAN}, a framework that leverages attention mechanisms within a GAN architecture to improve unsupervised image translation. Attention mechanisms allow the model to focus on significant regions of the input images, improving the quality of the generated output by emphasizing relevant features while ignoring less important details. This model employs a dual-branch architecture, where one branch is dedicated to the source domain and the other to the target domain.

They also propose AttentionGAN in \cite{tang2021attentiongan}, a framework that leverages attention mechanisms within a GAN architecture to facilitate unpaired I2I translation. AttentionGAN consists of a generator and a discriminator, where the generator utilizes attention modules to refine the image translation process. The incorporation of attention mechanisms that allow the model to focus on relevant features and regions of the input images enhances the model's ability to generate high-quality outputs by selectively emphasizing important details while minimizing distractions from irrelevant areas. This model achieves high-quality translations while maintaining semantic coherence.


Mejjati et al. in \cite{mejjati2018unsupervised} propose a framework that utilizes attention mechanisms to perform I2I translation. The proposed model employs a dual-branch architecture that consists of separate pathways for different domains. This structure helps to capture domain-specific features while allowing for effective interaction between the two branches during the translation process. By focusing on relevant features and employing a dual-branch architecture, the model achieves high-quality translations while maintaining semantic consistency. This model integrates adversarial and cycle-consistency losses for both source and target domains, promoting semantic consistency between input and output images while ensuring the generated images exhibit the desired features of the target domain.

Kim et al. in \cite{kim2019u} propose U-GAT-IT, a framework for unsupervised image translation that incorporates attention mechanisms and adaptive normalization to improve the quality of generated images. The introduction of Adaptive Layer-Instance Normalization (ALIN), which combines layer normalization and instance normalization, allows the model to adaptively adjust the normalization process based on the characteristics of the input data. This adaptability enhances the quality of the generated images by better preserving style and content. By focusing on relevant features and specific regions of the input image and dynamically adjusting normalization processes, U-GAT-IT achieves high-quality translations. This model incorporates multiple loss functions including adversarial loss, cycle loss, and identity loss, which collectively ensure that the generated images are both visually appealing and semantically consistent with the target domain. There is also a light version of U-GAT-IT, called U-GAT-IT-light. U-GAT-IT-light is designed to be less demanding in terms of memory and processing power, enabling faster training and inference times, while maintaining effective image translation capabilities.

Zhan et al. in \cite{zhan2022bi} propose RABIT, a system for image translation and manipulation that is based on a Bi-level Feature Alignment Scheme with a Ranking and Attention mechanism. An approach to image translation that focuses on aligning features at two different levels: low-level features that capture fine details (e.g., textures and edges) and high-level features that capture more abstract semantic information (e.g., object shapes and semantics). By aligning both low-level and high-level features, the method preserves important details and semantic content. RABIT includes an alignment network and a generation network that work together. The alignment network uses an attention mechanism to match a conditional input with an exemplar, warping the exemplar to fit the input. The generation network then produces the final output using the warped exemplar and the conditional input. RABIT has mostly been used for image translation tasks, which use an example image as style guidance. The two-level alignment approach considerably reduces the computations and helps to build correspondences even with very high-resolution images.


The paper MixerGAN \cite{cazenavette2021mixergan} introduces a novel approach to image synthesis by effectively utilizing attention-based methods to retain long-range connections among image patches. Traditional convolutional networks often struggle with capturing these dependencies due to their limited receptive fields of convolutions, which can result in a loss of crucial contextual information. MixerGAN addresses this challenge by leveraging the MLP-Mixer architecture, which allows for efficient processing of image patches while maintaining global context without the high memory demands of transformer models. The architecture projects image patches into vectors processed through mixer blocks \cite{tolstikhin2021mlp}, facilitating interactions between distant pixels and enabling the model to account for global relationships that standard convolutional layers overlook.

The CWT-GAN \cite{lai2021unsupervised} model introduces an innovative attention-based approach that significantly enhances image-to-image translation by effectively utilizing weights from the discriminator in the generator's architecture. This cross-model weight transfer mechanism allows the generator to leverage the more expressive features learned by the discriminator, which is crucial for generating high-quality and diverse images. The model incorporates a residual attention mechanism that promotes the spread of features in the encoders, ensuring that important details are preserved during the translation process. By transferring weights from the discriminator's encoding module to the generator after each update, CWT-GAN capitalizes on the discriminator's ability to discern real images, thereby improving the generator's performance in producing realistic outputs. This attention mechanism is particularly beneficial in scenarios with significant shape differences between source and target domains. The integration of class activation maps (CAM) further enhances the model's capability to focus on critical areas of the image, ensuring that the generated outputs maintain semantic consistency and detail.

\subsection{VAE based Models}
VAE-based models (Variational Autoencoder) are used in image-to-image translation applications to learn the latent features of images and generate new images with similar characteristics. These models consist of two main parts: the encoder, which transforms the input into a latent space with a probabilistic distribution, and the decoder, which uses the latent features to generate a new image.

UNIT\cite{liu2017unsupervised} uses a shared latent space between two different image domains, such that each pair of images from the two domains can be mapped to a common latent code that reconstructs both images. This model is a combination of Variational Autoencoders (VAEs) and Generative Adversarial Networks (GANs). VAEs are responsible for encoding and decoding images within each domain, while GANs ensure the realism of the generated images through adversarial training. UNIT implicitly uses a shared latent space that structurally ensures consistency.

TransGaGa \cite{wu2019transgaga} presents a geometry-aware framework for unsupervised image-to-image translation, designed to handle significant shape variations between domains. It separates images into geometry and appearance spaces, enabling more precise transformations by disentangling structural and appearance components. This separation allows the model to focus on translating geometry without being affected by appearance changes, preserving the original structure. The framework, using auto-encoders and transformers, employs a conditional variational autoencoder for unsupervised learning and incorporates prior losses to enhance geometry estimation.


The MUNIT\cite{huang2018multimodal} auto-encoder architecture consists of three main components: the content encoder, the style encoder, and the decoder. The content encoder is made up of multiple strided convolutional layers followed by some residual blocks, which help extract complex features and reduce the dimensionality of the input image. In contrast, the style encoder consists of strided convolutional layers, a global average pooling layer, and a fully connected layer to generate style codes. The decoder then utilizes a Multi-Layer Perceptron (MLP) to generate Adaptive Instance Normalization (AdaIN) parameters from the style code. Subsequently, the content code undergoes processing through residual blocks equipped with AdaIN layers, before being reconstructed into image space through upsampling and convolutional layers. The bidirectional loss function includes both image reconstruction and adversarial loss, aiding in matching the distribution of generated images to that of real images. Furthermore, the nonlinearity of the decoder allows for a multimodal output image distribution. Finally, the concept of style-augmented cycle consistency ensures that translating back produces the original image while preserving the style.

The COCO-FUNIT\cite{saito2020coco} model is an encoder-decoder architecture for few-shot, unsupervised image-to-image translation. It extracts structural features from the content image and style information from the style image to generate a style code. The Constant Style Bias (CSB) emphasizes texture and color details, allowing better control over style integration. Adaptive Instance Normalization (AdaIN)\cite{huang2017arbitrary} merges the content features with the style code, preserving the original content while applying the desired style. This design enables realistic translations with minimal labeled data, improving few-shot translation performance.

Han et al. in \cite{EM-LAST} propose EM-LAST. This method utilizes multi-dimensional latent space to represent various aspects and features of images from different domains. EM-LAST utilizes an energy-based modeling approach to model the data distributions of both the source and target domains. In addition, it uses a transport mechanism that efficiently moves latent representations from the source domain to the target domain. This mechanism designed to preserve important semantic information, leading to high-quality image translations. Their results show that the combination of multi-dimensional latent space representation and energy-based modeling is enhancing the flexibility and robustness of generative models.

\subsection{Diffusion Models}

Diffusion models have gained attention for their potential in various generative tasks, including image-to-image translation. These models offer a probabilistic framework that can produce high-quality outputs by modeling the transformation of data through a series of stochastic processes.

Sasaki et al. in \cite{sasaki2021unit} propose UNIT-DDPM, a framework for unpaired image translation leveraging Denoising Diffusion Probabilistic Models (DDPM). DDPMs are a class of generative models that iteratively refine samples from noise to data distributions. UNIT-DDPM employs a two-step process where initial noisy images are progressively refined into target domain images using a learned diffusion process. The model captures the underlying structure of the data distributions in both source and target domains, facilitating effective translations. The incorporating of cycle consistency in this method helps maintain the semantic integrity of the images. Their results show that DDPMs can effectively capture complex image characteristics, leading to more realistic translations.

Zhao et al. in \cite{zhao2022egsde} propose EGSDE, a diffusion-based model for unpaired I2I translation. This framework utilize stochastic differential equations (SDEs) to define a continuous-time diffusion process that iteratively refines samples from noise to produce coherent images in the target domain. The core innovation of EGSDE lies in its use of energy-based models to guide the translation process. The framework formulates the image translation task as a stochastic process where energy landscapes help navigate the transformations between different image domains. The energy-guided approach helps in enabling the model to generate high-quality translations between different domains without requiring paired training data. By leveraging the principles of diffusion and energy-based methods, EGSDE aims to improve the quality and robustness of image translations.

Meng et al. in \cite{meng2022sdedit} propose SDEdit, a framework that utilizes stochastic differential equations (SDEs), based on a diffusion model generative prior, for both image synthesis and editing tasks. Users can specify desired attributes or modifications, allowing for targeted editing and synthesis of images based on specific intentions. Given an input image with user guide, SDEdit first adds noise to the input, and then subsequently denoises the resulting image through the SDE prior to increase its realism. This model can be integrated with existing pre-trained diffusion models, enhancing its efficiency and output quality.


Kim et al. in \cite{kim2023unpaired} propose UNSB, an approach to unpaired I2I translation that utilizes the concept of the Schrödinger bridge within a neural network framework. They introduce the Neural Schrödinger Bridge (NSB) as a way to model the optimal transport problem between two distributions of source and target domains in a probabilistic manner. This approach formulates the image translation task as an optimal transport problem with a regularization term. This allows the model to find a smooth and coherent mapping from the source domain to the target domain while maintaining the integrity of the data distributions. Their results show the effectiveness of the proposed method in generating realistic and coherent images.

\subsection{Flow based Models}
Flow-Based Generative Models are powerful methods for modeling data distributions, which are designed based on reversible and continuous transformations. These models use a series of simple and reversible mathematical mappings to transform complex data into a simple Gaussian distribution and vice versa. Their main feature is the ability to directly calculate the probability density of the data and accurately reproduce the samples.

The StyleFlow model \cite{fan2022styleflow} offers an image-to-image translation approach that preserves content while altering style, utilizing Normalizing Flows for invertible mappings and accurate density estimation. Based on the GLOW model \cite{kingma2018glow}, it transforms complex data distributions into simpler ones, effectively enabling content-preserving translation. The architecture includes a Style-Aware Normalization (SAN) module, invertible blocks, and a pre-trained VGG-19 encoder \cite{simonyan2014very} for feature extraction. The loss function comprises Content Loss (for content preservation), Aligned-Style Loss (to match the target style), and Smoothness Loss (to prevent artifacts and ensure smooth transformations). These components allow StyleFlow to perform high-fidelity, flexible style transformations in image-to-image translation.

In Hierarchy Flow \cite{fan2023hierarchy}, the Hierarchical Coupling Layer combines features from different network layers while preserving spatial integrity, eliminating the need for spatial downscaling. This allows for multi-scale feature extraction, maintaining fine details and structures. For style transfer, the model uses AdaIN \cite{huang2017arbitrary} to align the target image's features with the reference style image. The Aligned-Style Loss, which balances content preservation and style transfer, compares input-output features for content and aligns statistical properties for style, ensuring effective style application without compromising the image's structure.

The DUAL-GLOW model \cite{sun2019dual} is designed for translating MRI images to PET images. It uses invertible flow-based functions to model the conditional distribution of PET images given MRI features, mapping both to latent spaces. The conditional probability of PET images is modeled as a Gaussian distribution, with mean and variance determined by MRI features. To reduce computational cost, the model uses a hierarchical structure and data splitting techniques. Additionally, side information such as age and gender is incorporated at higher levels, allowing the model to generate PET images that retain MRI information while considering these factors.

\subsection{Transformers Models}
The integration of Transformer architecture into image-to-image (I2I) translation networks has introduced a robust framework for enhancing both global and instance-level understanding within images. Recent studies demonstrate the effectiveness of Transformers in unpaired I2I translation tasks by leveraging their powerful attention mechanisms and token-mixing capabilities.

In~\cite{zheng2022ittr}, ITTR is proposed as a Transformer-based model for unpaired I2I translation, featuring two core innovations: the Hybrid Perception Block (HPB) and Dual Pruned Self-Attention (DPSA). The HPB integrates local and global perception by employing a dual-branch design—one branch uses depth-wise convolution for efficient local token mixing, while the other utilizes self-attention for capturing long-range dependencies. A multi-layer perceptron (MLP) fuses the outputs of these branches, ensuring effective contextual integration. The DPSA mechanism reduces computational overhead by ranking and pruning less-contributive tokens before calculating the attention map, making the attention process more efficient without compromising performance. ITTR’s architecture, composed of overlapping patch embedding, nine stacked HPBs, and a symmetric decoder, achieves state-of-the-art performance on six benchmark datasets, demonstrating both efficiency and effectiveness in unpaired I2I translation tasks.

InstaFormer~\cite{kim2022instaformer} is an instance-aware image-to-image translation architecture based on transformers, seamlessly integrating both instance-level and global information. Unlike conventional methods, InstaFormer uses a self-attention module in Transformers to discover global consensus across content features while augmenting these features with instance-level information, extracted from object bounding boxes, to enhance the translation of object regions. The framework replaces standard layer normalization with adaptive instance normalization (AdaIN) for multi-modal translation, enabling the use of style codes. It also introduces a novel instance-level content contrastive loss to improve translation quality, particularly in object regions, and employs a one-sided framework to avoid the distortions typically introduced by cycle-consistency in other I2I models.

UVCGAN~\cite{torbunov2023uvcgan} is an enhanced version of CycleGAN for unpaired image-to-image translation, which incorporates a UNet-ViT generator to improve performance while maintaining cycle consistency. The model utilizes a hybrid architecture with a Vision Transformer (ViT) at the bottleneck, allowing it to effectively learn relationships between low-frequency features. To further improve results, the discriminator loss is augmented with a gradient penalty (GP), and the generators are pre-trained in a self-supervised manner on an image inpainting task. These innovations lead to better performance in preserving correlations between the original and translated images compared to existing models.

\section{Datasets} \label{data_sec}

\begin{figure}
\centering
\subfloat[Aerial photo to map]{\includegraphics[width=0.4\textwidth]{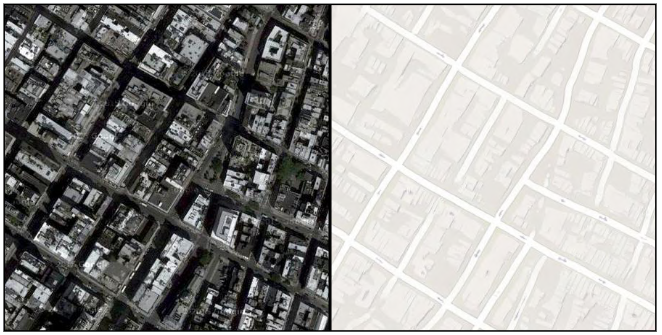}\label{Paired_2}}\hskip1ex
\subfloat[zebra to horse ]{\includegraphics[width=0.4\textwidth]{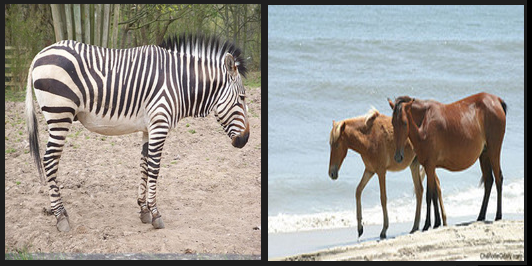}\label{unPaired_2}}



\caption{(a) and (b) are paired datasets. (c) and (d) are unpaired datasets. }
\end{figure}

Image-to-image translation is a significant problem in computer vision and machine learning that deals with the learning of the mapping between an input image and an output image for applications like style transfer, data augmentation, and image retrieval. The process involves aligning input images from the source domain with the target domain to generate the desired output. Paired datasets, which consist of corresponding input-output image pairs, are ideal for supervised models such as Pix2Pix. These models achieve high accuracy but require time-consuming and costly data preparation \ref{Paired_2}. In contrast, unpaired datasets, where input and output images are independent, are more suitable for unsupervised models like CycleGAN \ref{unPaired_2}. While these models offer greater flexibility and can be applied to a wider range of tasks, they typically have lower accuracy than those based on paired datasets.

Several benchmark datasets are available that can be used to perform Image-to-Image Translation tasks. Table \ref{Dataset} lists the dataset names, references, and a brief description of each dataset.
Then, in Table \ref{task} the datasets from the previous table are used to create dual datasets and define different tasks for image-to-image translation.

One of the most important criteria in image-to-image translation is content preservation, which means that the features and information of the source image should be clearly identifiable in the converted image. Based on the source domain, target domain, and degree of content preservation, data are classified into three main categories: full content preservation (FCP), partial content preservation (PCP), and non-content preservation (NCP). Each category is explained below, and Table \ref{task} shows the type of each dataset constructed. This classification helps us to better select the type of problem and the appropriate method for image-to-image translation. This topic has been less discussed in other papers. The aim of this section is to review the existing data with respect to content preservation.

\begin{table}[htbp]
  \centering
  \caption{Dataset and description}
    \scriptsize
    \begin{longtable}{|p{2cm}|p{0.8cm}|p{12cm}|}
    \hline
    dataset & ref.  & Description \\
    \hline
    
     Cityscape &
     \cite{cordts2016cityscapes}
     & Contains a diverse set of stereo video sequences recorded in street scenes from 50 different cities, with pixel-level annotations on the frames \\
     \hline
 ADE20k & 
 \cite{zhou2017scene}
 & Spans diverse annotations of scenes, objects, parts of objects, and even parts of parts. Includes 150 object and stuff classes, serving as the foundation for a scene parsing benchmark \\
 \hline
 AFHQ  & 
 \cite{choi2020stargan}
 & Consists of 15,000 high-quality 512 × 512 images of animal faces across three domains: cats, dogs, and wildlife, each with 5,000 images \\
 \hline
 Alderley & 
  \cite{milford2012seqslam}
 & Includes GPS-tagged images captured on the same route twice: once on a sunny day and once on a stormy night, allowing for corresponding day-night frame pairs \\
 \hline
 BDD   & 
 \cite{yu1805bdd100k}
 & Large-scale, high-resolution autonomous driving dataset with 100,000 video clips from various cities and conditions, featuring key frames annotated with bounding boxes and dense pixel details. \\
 \hline
 CelebA & 
  \cite{liu2015deep}
 & Contains 200,000 images of 10,000 identities (20 images per identity), ideal for training face recognition models \\
 \hline
 Deepfashion & 
 \cite{liu2016deepfashion}
 & Consists of 52,712 images of people wearing fashion items \\
 \hline
 estate & 
  \cite{poursaeed2018vision}
 & Combines images from LSUN, Places, and Houzz, with around one million labeled images for scene and object categories \\
 \hline
 Facades & 
 \cite{tylevcek2013spatial}
 & Combines ECP-Monge and eTrims datasets, with images annotated in eight classes and a large view dataset. \\
 \hline
 FFHQ  & 
  \cite{karras2019style}
 & Flickr-Faces-HQ is a high-quality human face image dataset, used as a benchmark for GANs. \\
 \hline
 Flickr & 
 \cite{plummer2015flickr30k}
 & Adding 244k coreference chains and 276k bounding boxes to 158k captions helps advance automatic image description and grounded language understanding \\
 \hline
 GTA   & 
  \cite{richter2016playing}
 & Contains 24,966 semantically labeled frames from GTA5, each with a resolution of 1914×1052 pixels \\
 \hline
 ImageNet & 
 \cite{deng2009imagenet}
 & Based on the WordNet hierarchy, ImageNet features over 100,000 synsets (mainly nouns), with an average of 1,000 quality-controlled images per synset, totaling tens of millions of labeled images \\
 \hline
 KITTI & 
  \cite{geiger2012we}
 & Captured with an autonomous driving platform using high-resolution cameras, a Velodyne laser scanner, and GPS, including images of up to 15 cars and 30 pedestrians in various settings \\
 \hline
 NYUv2 & 
 \cite{silberman2012indoor}
 & Includes video sequences of indoor scenes recorded with RGB and Depth cameras from Microsoft Kinect, with each object labeled by class and instance number \\
 \hline
 OpenPose & 
  \cite{cao1812openpose}
 & Supports a real-time system that detects 135 keypoints for human bodies, feet, hands, and faces in single images \\
 \hline
 Satellite-Googlemaps-Masks &
  \cite{isola2017image}
 & Consists of 2,107 triplets (600x600), including satellite images, corresponding Google Map snapshots, and binary masks separating roads from other terrains. \\
 \hline
 SUNCG & 
 \cite{bi2019deep}
 & Aims to create a full 3D voxel representation and semantic labels from a single-depth view, using a large-scale, manually annotated dataset of synthetic 3D scenes \\
 \hline
 UT-Zap50K & 
  \cite{yu2014fine}
 & Consists of 50,025 catalog images from Zappos.com, categorized into shoes, sandals, slippers, and boots, with further subdivisions by function and brand \\
 \hline
 wikiart & 
  \cite{kalkowski2015real}
 & Online database of visual arts that includes artworks, artist information, and descriptions related to the works \\
 \hline
 YFCC100M & 
 \cite{kalkowski2015real}
 & Contains 100 million media objects, including 99.2 million photos and 0.8 million videos, each with metadata like Flickr identifier, owner name, camera details, title, tags, and geographic data \\\hline
 
 U-GAT-IT &
 \cite{kim2019u}
 & Contains 46,836 selfie images annotated with 36 attributes and 69,926 anime character images from Anime-Planet \\\hline
    \end{longtable}%
  \label{Dataset}
\end{table}%

\subsection{Fully Content preserving(FCP)}

\begin{figure}
\centering
\subfloat[GTA data to real city data]{\includegraphics[width=0.4\textwidth]{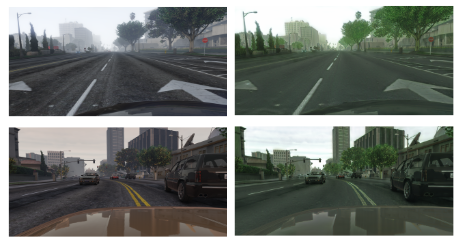}\label{gta2cs}}\hskip1ex
\subfloat[paintings to real images.]{\includegraphics[width=0.4\textwidth]{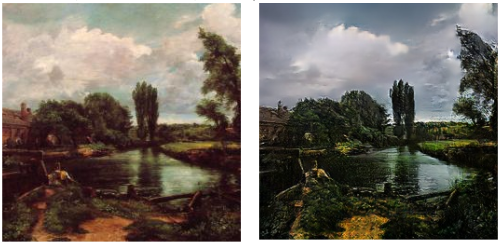}\label{art2real}}

\vspace{0.1cm}

\subfloat[Cityscapes to segmentation labels]{\includegraphics[width=0.45\textwidth]{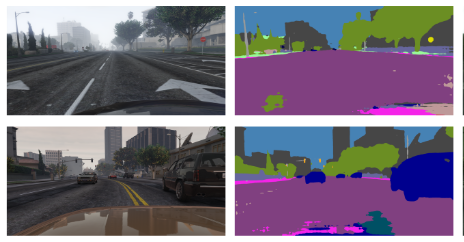}\label{20}}\hskip1ex
\subfloat[Horse to Zebra]{\includegraphics[width=0.45\textwidth]{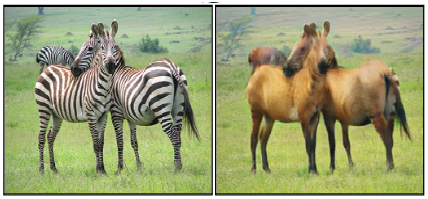}\label{4}}

\vspace{0.1cm}

\subfloat[Selfie to Anime]{\includegraphics[width=0.45\textwidth]{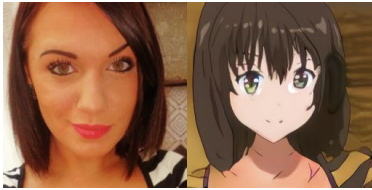}\label{NCP1}}\hskip1ex
\subfloat[Cat to Dog]{\includegraphics[width=0.45\textwidth]{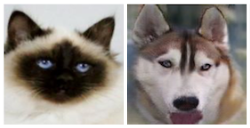}\label{NCP2}}

\caption{(a) and (b) are Fully Content preserving . (c) and (d) are Partially Content preserving . (e) and (f) are Non Content preserving}
\end{figure}

In tasks where complete content preservation is crucial, the model must accurately reproduce all the details and information present in the input image in the output image. This means that the translated images should retain the content of the source images while aligning the data distribution with the target images. Fully preserving content is essential for applications such as image enhancement (increasing resolution, brightness, contrast, and other visual features), converting simulated images to real ones, and style transfer, as any loss of detail can lead to incorrect results or loss of information.
In Fig 
\ref{gta2cs}
and 
\ref{art2real}
Examples of translating simulated GTA images into real images and converting painted images into real images are shown.

%
%
%
%
%

\subsection{Partially Content preserving(PCP)}


In tasks where only certain content needs preservation, the model should retain essential sections of the input image while allowing changes in other areas. Despite differences in visual domains of the source and target, they share structural information, aiming to transfer the source domain's visual characteristics to the target. Partial content preservation is common in applications like seasonal translation and noise removal, such as maintaining a horse's body structure while altering skin color and patterns in a horse-to-zebra conversion. In Fig \ref{20} and \ref{4} examples of translate Cityscapes into segmentation labels and horse to zebra are shown.

\subsection{Non Content preserving(NCP)}

In image-to-image translation tasks where content preservation is not required, images from the source and target domains may be quite different in appearance. The translation process in such cases occurs at the semantic level, meaning that the content of the source image is allowed to undergo significant changes in order to match the features of the target domain. In these scenarios, the focus is less on preserving specific content from the source and more on ensuring that the transformed image conveys the correct meaning or visual properties relevant to the target domain.
Such transformations are often used in creative and artistic projects, where the goal is to produce entirely new visual representations that may not resemble the original images in terms of style, appearance, or even object details.
In Fig.
\ref{NCP1}
and
\ref{NCP2}.

	



\begin{table}[htbp]
  \centering
  \caption{Task: Defines the source domain and target domain for the translation process,
Dataset: Indicates the name of the dataset from which the source and target domains were selected,
Category: Specifies the type of content retention involved in the translation}
    \resizebox{\textwidth}{!}{
    \begin{tabular}{|c|c|c|}
    \hline
    Task   & Dataset
    & Category \\
    \hline

    Cezanne2Real(Cz2R) \cite{tomei2019art2real} & wikiart \& Flickr & FCP \\

    CG2Real(CG2R)  \cite{liu2021separating} & SUNCG \& (estate+NYUv2) & FCP \\

    GTA2Cityscape(G2C)  \cite{ozaydin2022dsi2i} & GTA \& Cityscape & FCP \\

    GTA2KITTI(G2K)  \cite{fan2022styleflow} & GTA \& KITTI & FCP \\

    Landscapes2Real(L2R)  \cite{tomei2019art2real} & wikiart \& Flickr & FCP \\

    Monet2Real(M2R)  \cite{tomei2019art2real} & wikiart \& Flickr & FCP \\

    Photo2Portrai\_CelebA\_Wikiart(PH2P)  \cite{zhao2020feature} & WikiArt \& CelebA & FCP \\

    Portraits2Real(P2R)  \cite{tomei2019art2real} & wikiart \& CelebA & FCP \\

    Ukiyo-e2Real(U2R) \cite{tomei2019art2real} & wikiart \& Flickr & FCP \\

    Vangogh2Photo(V2PH) \cite{tomei2019art2real} & wikiart \& CelebA & FCP \\
\hline

    Aerial2Map(A2M) \cite{liu2017unsupervised}  & Satellite-Googlemaps-Masks & PCP \\

    Anime2Selfie(An2S) \cite{lee2023progressively} & U-GAT-IT* & PCP \\

    Apple2Orange(Ap2O) \cite{lee2023progressively} & ImageNet & PCP \\

    Edges2Shoes(E2SH) \cite{lin2019exploring} & UT-Zap50K & PCP \\

    Horse2Zebra(Sm2W) \cite{lee2023progressively} & ImageNet \& ImageNet & PCP \\

    Label2Cityscape(L2C) \cite{CUT}   & Cityscapes & PCP \\

    Label2Photo(L2PH) \cite{zhan2022bi} & ADE20k & PCP \\

    Labels2Facades(L2Fa) \cite{sasaki2021unit} & Facades & PCP \\

    Night2Day\_Alderley(N2D\_A) \cite{zheng2020forkgan} & Alderley & PCP \\

    Night2Day\_BDD(N2D\_B) \cite{zheng2020forkgan} & BDD   & PCP \\

    summer2winter(Sm2W) \cite{lee2023progressively} & Flickr \& Flickr & PCP \\
\hline
    Cat2Dog(C2D) \cite{lee2023progressively} & ImageNet \&ImageNet & NCP \\

    Cat2Dog\_AFHQ(C2D\_AF) \cite{CUT}   & AFHQ  & NCP \\

    CelebA-HQ(Ce) \cite{cao2020informative} & CelebA-HQ & NCP \\

    Glass2No-Glass(G2NG) \cite{zhao2020unpaired} & CelebA & NCP \\

    Human2Cat(H2C) \cite{wu2019transgaga} & CelebA \& YFCC100M & NCP \\

    Keypoint2Photo(K2PH) \cite{katzir2019cross} & OpenPose \& Deepfashion & NCP \\

    Lion\_sealion2Tigert\_tigerbeetle(Li2Ti) \cite{xie2021unaligned} & Imagenet & NCP \\

    Male2Female(ML2FL)\cite{zhao2022egsde}       &  CelebA-HQ & NCP \\

    man2woman(M2W) \cite{shubhra2018iegan} & FFHQ  & NCP \\

    Wild2Dog(Wi2D) \cite{zhao2022egsde} & AFHQ  & NCP \\
    \hline

    \end{tabular}
    }%
  \label{task}%
\end{table}%

\section{Different Metrics} \label{Metrics_sec} 
In this section, we introduce criteria for evaluating image production and image-to-image translation. 
The $IS$ metric \cite{salimans2016improved} uses a pre-trained Inception network to compute the KL divergence between the conditional label distribution $p(y|x)$ and the marginal distribution $p(y)$. This metric evaluates the quality and diversity of the generated images. The higher the $IS$ value, the better the quality and diversity of the synthetic images. In equation \ref{ISeq}, we have shown the formula for calculating $IS$.

\begin{equation}
			IS = exp(\mathbb{E}_x[KL(p(y|x)||p(y))])
			\label{ISeq}
        \end{equation}
 
 Where $x$ represents the sample images (either synthetic or real) and $p(y|x)$ is the label distribution obtained from the Inception Network's classification. The upper bound of $IS$ is determined for real data, and the closer to upper bound of the $IS$ means the synthetic images have better  quality and diversity.

The FID measures the similarity between two sets of images, typically comparing generated images to real images. Lower FID scores indicate that the generated images are more similar to the real images, suggesting higher quality and more realistic outputs. FID measures the distance between the feature vectors of real and generated images. The features of these two sets of images are extracted using the Inception network~\cite{szegedy2015going}. A lower FID score indicates greater similarity between the two groups of images, with a score of 0 representing identical images. FID is commonly used to evaluate the quality of images generated by GANs, with lower scores highly correlated with higher quality images.The mean of class-wise Fréchet Inception Distance (mFID) indicates performance quality: lower scores signify better performance, contrasting with Inception Score (IS) evaluations.
In relation to \ref{FIDeq}, the method of calculating the FID value for two sets of real data $r$ and synthetic data $g$ is shown \begin{equation}
        	FID = {||\mu_r - \mu_g||}_2^2 + Tr(\Sigma_r + \Sigma_g - 2(\Sigma_r\Sigma_g)^{1/2})
        	\label{FIDeq}
        \end{equation}
Where $\mu_r$ and $\mu_g$ are the mean feature vectors for the real and synthetic data, respectively, and $\Sigma_r$ and $\Sigma_g$ represent the covariance of the two sets. A lower $FID$ value indicates a greater similarity between the generated and real images.

The Single Image FID (SIFID) metric is another metric based on FID to compare real and generated distributions. Instead of relying on the activation vector obtained from the final pooling layer of the Inception Network (which provides a single vector per image), the internal distribution of deep features at the output of the convolutional layer immediately preceding the second pooling layer is utilized (one vector per location in the map)\cite{shaham2019singan}.

KID is similar to the FID metric, but it does not assume a normal distribution like FID~\cite{binkowski2018demystifying}. While FID assumes that the extracted features from the Inception network follow a normal distribution, KID does not make this assumption. Instead, KID directly measures the dissimilarity between the distributions of real and generated images using the Maximum Mean Discrepancy (MMD). In essence, KID computes the squared MMD between the feature representations of real and generated images. This approach allows KID to capture differences between distributions without relying on specific distributional assumptions, providing a potentially more robust evaluation metric in generation tasks.

Both FID and KID work in the low mode such that that lower values indicate better performance. This is because both FID and KID metrics measure the dissimilarity between the distributions of real and generated images. 

Comparing to KID, fused-KID is beneficial for scenery tasks because it considers both foreground and background elements~\cite{lin2021attention}. In addition to its comprehensive evaluation of foreground-background relationships, fused-KID provides a nuanced analysis of contextual interactions within scenic imagery.

The IS~\cite{salimans2016improved} evaluates the quality and diversity of the generated images. It uses a pre-trained Inception network to classify the images and measures the entropy of the predicted class distribution. Higher IS scores indicate that the generated images are both diverse and high quality.

The Learned Perceptual Image Patch Similarity (LPIPS)~\cite{zhang2018unreasonable} metric assesses how similar two images are from a perceptual view. It does this by comparing the activations of corresponding patches in the images within a pre-trained neural network. This method has been found to align closely with human visual perception. A lower LPIPS score indicates that the image patches are perceptually similar. LPIPS calculates the mean distance between the features of generated samples. A higher LPIPS score means that the output images are more diverse.

The Number of Statistically-Different Bins (NDB) and Jensen–Shannon Divergence (JSD)~\cite{richardson2018gans} are two bin-based evaluation metrics used to assess the similarities between generated distributions and real distributions. However, relying solely on diversity metrics does not help guarantee that the distribution of generated images matches with that of real data. To address this, two recently proposed bin-based metrics \cite{richardson2018gans} are incorporated. The NDB assesses the relative proportions of samples that have been categorized into clusters based on real data, and JSD, which quantifies the similarity between distribution bins.

Mean Match Rate (MMR), Mean Object Detection Score (MODS), and Mean Valid IoU Score (MVIS)~\cite{su2021towards}. MVIS specifically measures the average Intersection over Union between the predicted masks and the corresponding generated masks. This metric is used when we want to convert image domain to mask domain in segmentation tasks.

Peak Signal-to-Noise Ratio (PSNR) is a widely used metric that measures the quality of reconstructed images compared to their original counterparts, with higher values indicating better quality and less distortion. Structural Similarity Index (SSIM) assesses the perceived quality of images by comparing luminance, contrast, and structure, providing a more comprehensive evaluation of visual similarity than PSNR~\cite{wang2004image}.

%
%

\section{Evaluation} \label{Evaluation_sec}

The experiments conducted for the studied models in their base papers are organized based on the categorization of tasks in section \ref{data_sec} and are presented in this section. We examine the results of many image-to-image translation models that we have studied in this paper, more than sixty models. The visual results and values of various metrics mentioned in Section \ref{Metrics_sec} and for the tasks and datasets introduced in Section \ref{data_sec} are reviewed. But due to the large volume of results, we have selected two metrics, FID and KID, and presented their results in the tables of this section. Because not all models have been tested with all the evaluation metrics in their base papers, this issue can also be seen in the presented tables in this section and some cells in the tables that are relevant to the mentioned issue are empty.

\subsection{Fully Content preserving}

In this section, we present the results of I2I models for tasks which are Fully Content preserving. In such tasks where full content preservation is intended, the model needs to reproduce with high accuracy all details and information from the input image into the output image. For the tasks that fall into this category, we have presented the results of the I2I models on two metrics in separate Tables. The names of the tasks in these Tables are written with abbreviations, but the necessary explanations for them are given in the captions of each table. 

\begin{figure}[tbp]
	\centering
	\includegraphics[width=0.95\textwidth]{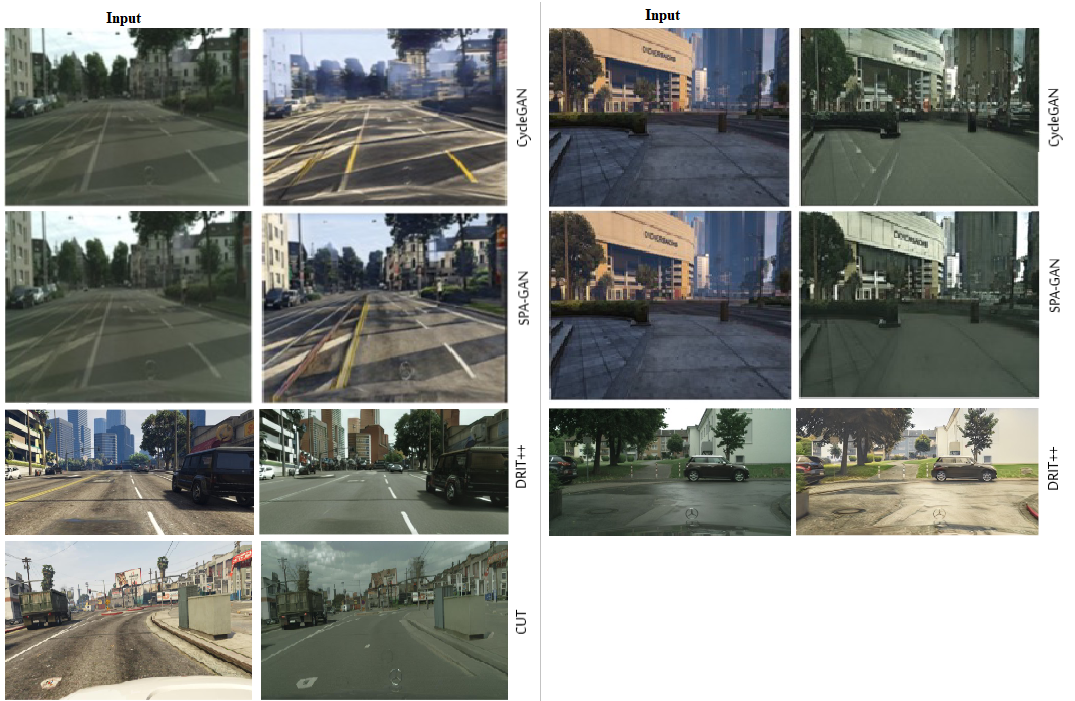}
	\caption{ 
	Visual comparison of I2I methods for GTA $\rightarrow{}$ Cityscape (left) and Cityscape $\rightarrow{}$ GTA (right) tasks. These results are taken from \cite{CUT, lee2018diverse, emami2020spa}.
	}
	\label{FCP_samples_fig} 
\end{figure}

In Fig. \ref{FCP_samples_fig}, we show some selected visual results for GTA $\rightarrow{}$ Cityscape and Cityscape $\rightarrow{}$ GTA tasks. The results of the I2I models for metrics FID and KID are given in Tables \ref{Fully_FID} and \ref{Fully_KID}, respectively. These results are taken from the models' own articles. From Table \ref{Fully_FID},  as an example, it can be found that the DRIT++ achieve the lowest FID among other methods for GTA $\rightarrow{}$ Cityscape and GTA $\rightarrow{}$ KITTI tasks. In Table \ref{Fully_KID}, the CUT method performs better than other methods on GTA $\rightarrow{}$ Cityscape task, based on KID metric.

\begin{table}[]	

	\centering
	\caption{FID values of different methods for Fully Content preserving category of tasks. \textbf{Abbreviations:} G: GTA, C: Cityscape, K: KITTI, R: Real, L: Landscapes, M: Monet, Cz: Cezanne, U: Ukiyo-e, P: Portraits, and CG: CG.}
\label{Fully_FID}	
    \hspace*{-1.8cm}
    \scalebox{0.75}{
	\begin{tabular}{|c|c|c|c|c|c|c|c|c|c|c|c|c|c|c|}
		\hline
		\multicolumn{1}{|c|}{Method/Task} & M2R & R2M & P2R & G2C & Cz2R & G2K & L2R & R2Cz & R2U & U2R & R2P & C2G & R2CG & CG2R \\ \hline
        		
        Art2Real \cite{tomei2019art2real} & 44.71 & - & 34.03 & - & 68  & - & 35.03 & - & -  & 80.48 & - &  - & -  & - \\ \hline
        CUT \cite{CUT}  & -      & - &  -     & 50.89 &   -    & 51.6  &    -   &   -    &   -    &   -    &   -    & 65.68 &   -    & - \\ \hline
        CycleGAN \cite{zhu2017unpaired} & 49.7  & 133.08 & 30.6  & 125.1 & 85.11 & 111.5 & 44.79 & 216.76 &  180.33 & 98.13 &  -     &     -  &    -   & -  \\ \hline
        DiscoGAN \cite{kim2017learning} &    -   & 279.41 &     -  &   -    &      - &     -  &     -  & 360.41 & 316.84 &     -  &     -  &  -     &  -      &  - \\ \hline
        DRIT \cite{lee2018diverse}  & 68.32 &  -     & 44.33 & 42.93 & 109.36 &  -     & 59.84 &  -     &    -   & 117.07 &    -   & 48.18 & 85.47 & 79.98 \\ \hline
        DRIT++ \cite{lee2018diverse} &     -  &   -    &  -     & 30.3  &   -    & 49.4  &    -   &   -    &    -   &    -   &   -    &      - &     -  & - \\ \hline
        DSMAP \cite{chang2020domain} & 81.61 & 63.94 & 62.44 &    -   &    -   &    -   &    -   & -       &    -   &     -  & 45.81 &     -  & 76.07 & 83.31 \\ \hline
        GDWCT \cite{GDWCT} & 113.16 & 71.68 & 83.69 &    -   &   -    &   -    &   -    &  -     &   -    &    -   & 75.86 &     -  &    -   & - \\ \hline
        HierarchyFlow \cite{fan2023hierarchy} & - & - & - &   70.87  &    -   &  -     &    -   &  -     &    -   &    -   &  &      - &     -  & - \\ \hline
        MSGAN \cite{mao2019mode} & 86.72 & 80.37 & 57.07 &     -  &    -   &  -     &    -   &  -     &    -   &    -   & 57.84 &      - &     -  & - \\ \hline
        MT-GAN \cite{lin2020learning} & 19.73 & 18.42 &   -    &    -   & 22.57 &   -    &   -    &  -     &   -    &  -     &  -     &      - &    -   & - \\ \hline
        MUNIT \cite{huang2018multimodal} & 85.06 & 77.85 & 93.45 & 47.76 &     -  & 375.2 &    -   & -       &    -   &   -    & 89.97 & 48.91 & 79.53 & 74.61 \\ \hline
        RF-GAN \cite{koksal2020rf} &    -   & 128.2 &    -   &    -   &    -   &   -    &   -    & 212.7 & 150.27 &    -   &    -   &   -    &   -    & - \\ \hline
        StarGAN \cite{choi2018stargan} & 35.95 & 172.02 &     -  &    -   & 31.37 &   -    & -      & 280.43 & 212.31 &     -  &   -    &   -    &  -     & - \\ \hline
        UNIT \cite{liu2017unsupervised} & 56.18 &   -    & 43.47 & 113.4 & 97.91 & 108.2 & 47.87 &     -  &     -  & 89.15 &   -    &      - &     -  & - \\ \hline
		
	\end{tabular} }
	\hspace*{-1cm}
\end{table}

\begin{table}[]	
	\centering
	\caption{KID values of different methods for Fully Content preserving category of tasks. \textbf{Abbreviations:} G: GTA, C: Cityscape, K: KITTI, PH2P: Photo2Portrai\_CelebA\_Wikiart and P2PH: Portrai2Photo\_CelebA\_Wikiart }
\label{Fully_KID}	
	\begin{tabular}{|c|c|c|c|c|}
		\hline
		\multicolumn{1}{|c|}{Method/Task} &  G2C & G2K & PH2P & P2PH    \\ \hline
		
        CUT \cite{CUT}  & 0.039 & 0.0832 &    -   & - \\ \hline
        CycleGAN \cite{zhu2017unpaired} & 0.0483 & 0.0837 & 1.84  & 1.82 \\ \hline
        DRIT \cite{lee2018diverse}  & 0.1469 &    -   &   -    & - \\ \hline
        DRIT++ \cite{lee2018diverse} & 0.1113 & 0.1382 &   -    & - \\ \hline
        FQ-GAN \cite{FQ-GAN} & 1.09  &    -   &    -   & 0.73 \\ \hline
        GcGAN \cite{fu2019geometry} & 0.1232 &    -   &    -   & - \\ \hline
        MUNIT \cite{huang2018multimodal} & 0.0868 & 0.1015 &  -     & - \\ \hline
        SRUNIT \cite{jia2021semantically} & 0.1627 &  -     &    -   & - \\ \hline
        StyleFlow \cite{fan2022styleflow} & 0.0432 & 0.0714 &   -    & - \\ \hline
        U-GAT-IT \cite{kim2019u} &    -   &     -  & 1.79  & 1.69 \\ \hline
        UNIT \cite{liu2017unsupervised}  & 0.0908 & 0.1352 & 1.2   & 1.42 \\ \hline
        VSAIT \cite{theiss2022unpaired} & 0.0874 &   -    &   -    & - \\ \hline
		
	\end{tabular}
\end{table}

\subsection{Partially Content preserving}
 
\begin{figure}
  \centering
  \begin{tabular}{@{}c@{}}
    \includegraphics[width=.9\linewidth,height=100pt]{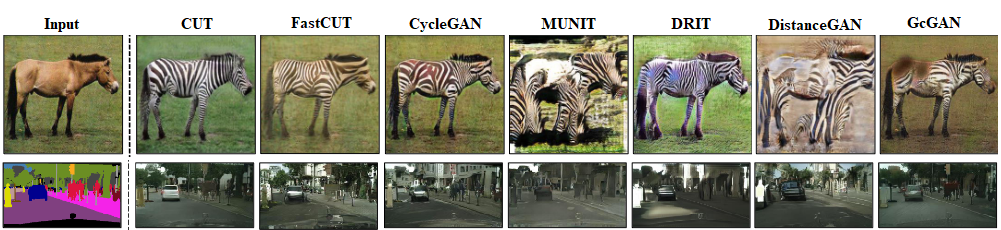} \\[\abovecaptionskip]
    \small (a) Results for Horse $\rightarrow{}$ Zebra (top) and Label $\rightarrow{}$ Cityscapes (bottom) tasks \cite{CUT}
  \end{tabular}

  \vspace{\floatsep}

  \begin{tabular}{@{}c@{}}
    \includegraphics[width=.8\linewidth,height=100pt]{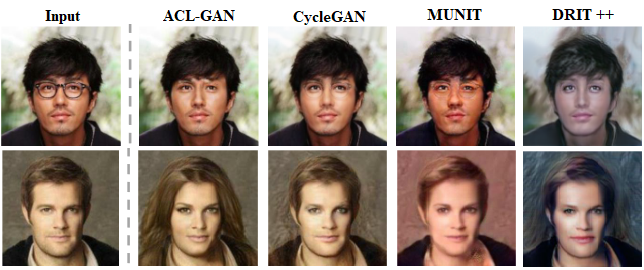} \\[\abovecaptionskip]
    \small (b) Results for Glass $\rightarrow{}$ No-Glass (top) and Male $\rightarrow{}$ Female (bottom) tasks \cite{zhao2020unpaired}
  \end{tabular}

  \caption{Visual results for some PCP and NCP tasks.}
  \label{PCP_NCP_samples_fig}
\end{figure}

In this section, we present the results of I2I tasks which are Partially Content preserving. In this category, the translation model should retain the necessary and significant parts of the input image, but allows changes in other details. Similar to the previous section for tasks in this category, we present the results of the I2I models on different metrics in separate Tables. In Fig.~\ref{PCP_NCP_samples_fig}(a), we present some results for Horse $\rightarrow{}$ Zebra and Label $\rightarrow{}$ Cityscapes tasks.
The results of the I2I models for metrics FID and KID, are given in Tables \ref{Partially_FID} and \ref{Partially_KID}, respectively. In Table \ref{Partially_FID}, we can see that for Horse $\rightarrow{}$ Zebra task, the UNSB has the lowest FID. As another example, DCLGAN performs better than other models on Label $\rightarrow{}$ Cityscapes task.

\begin{table}[]	
	\centering
	\caption{FID values of different methods for Partially Content preserving category of tasks. \textbf{Abbreviations:} L: Label, C: Cityscape, A: Aerial, M:Map, S: Selfie, An: Anime, F: Face, Ap: Apple, O: Orange,  E: Edges, H: Handbags, SH: Shoes, Fa: Facades, PH: Photos, V: Vangogh, Sm: Summer, W: Winter, Ho: Horse, Z: Zebra, N2D\_B: Night2Day\_BDD and  N2D\_A: Night2Day\_Alderley  }
\label{Partially_FID}	
	\hspace*{-2.5cm}
	\scalebox{0.65}{
	\begin{tabular}{|c|c|c|c|c|c|c|c|c|c|c|c|c|c|c|c|c|}
		\hline
		\multicolumn{1}{|c|}{Method/Task} & Ho2Z & Sm2W & W2Sm & L2C & L2PH & L2Fa & PH2V & O2Ap & S2An & M2A & E2SH & Ap2O & V2PH & N2D\_B & N2D\_A & A2M  \\ \hline 
        		
        CUT \cite{CUT}  & 45.5  & 84.3  &  -     & 56.4  & 56.4  & 119.7 & 96.9  & 127   & 87.2  &    -   & -      &  -     &    -   &   -    &     -  & - \\ \hline
        CWT-GAN \cite{lai2021unsupervised} & 85.44 & 74.92 & 76.99 &  -     &      - &     -  &     -  &   -    &      - & -    &  -     &   -    &    -   &   -    &      - & - \\ \hline
        CycleGAN \cite{zhu2017unpaired} & 77.2  & 78.76 & 79.58 & 76.3  & 76.3  & 127.5 & 119.85 & 117.7 & 99.69 & 216.89 & 75.54 & 181.6 & 85.1  & 51.7  & 167   & 150.23 \\ \hline
        CycleGAN-MGGC \cite{ganjdanesh2024compressing} &  55.06 &  75.85 &  -  &  - & -  & - & - & -  & -  &  -   &  - &  -  &    -   &   -    &     -  & - \\ \hline
        DCLGAN \cite{han2021dual} & 43.2  &   -    &   -    & 49.4  & 49.4  & 119.2 & 93.5  & 124.9 &  -     &  -     & -      &   -    & -      &     -  &  -     & - \\ \hline
        DistanceGAN \cite{benaim2017one} & 72    & 97.2  &  -     & 81.8  & 81.8  &    -   &      - &   -    &  -    &      - &     -  &  -     & -      &    -   &     -  & - \\ \hline
        DivCo \cite{liu2021divco} &    -   & 51.57 & 45.79 &     -  &    -   & 75.96 &   -    &  -     &  -     & 109.54 &      - & -      &   -    & -      &  -     & - \\ \hline
        DRIT \cite{lee2018diverse}  & 140   & 81.64 & 41.34 & 155.3 & 155.3 & -      & 136.24 & 141.07 & 104.49 &      - & 84.64 & 176.13 & 108.92 & 53.1  & 145   & - \\ \hline
        DRIT++ \cite{lee2018diverse} & 88.5  & 258.5 & 93.1  & 96.2  &    -   & 336.76 &    -   &    -   & 104.4 & 316.42 & 123.87 &       &  -     &   -    & -      & 237.94 \\ \hline
        FastCUT \cite{CUT} & 73.4  &  -     &  -     & 68.8  & 68.8  &  -     &    -   &   -    &     -  &   -    &    -   &   -    &    -   &   -    &   -    & - \\ \hline
        F-LSeSim \cite{zheng2021spatially} & 38    & 224.9 &   -    & 49.7  &      - &   -    &    -   &  -     &     -  &  -     &   -    &   -    &   -    &   -    &    -   & - \\ \hline
        FSeSim \cite{zheng2021spatially} & 40.4  &    -   & 90.5  & 58.1  & 53.6  &     -  &    -   &    -   & -      & -      & -      &  -     &    -   &     -  &  -     & - \\ \hline
        GcGAN \cite{fu2019geometry} & 86.7  & 97.5  & -      & 57.4  & 105.2 &     -  &     -  &     -  & 267.73 &  -     &   -    & -      &    -   &     -  &   -    & - \\ \hline
        IEGAN \cite{ye2021independent} &     -  & 97.71 & 83.35 &     -  &     -  &     -  &     -  & 132.86 &   -    &    -   &     -  & 164.58 &   -    &    -   &   -    & - \\ \hline
        InvolutionGAN \cite{deng2023involutiongan} & 62.929  & - & - &  -  &   -  &    -  & 106.499  & - &   72.033   &  -   &   -  & - &   -    &    -   &   -    & - \\ \hline
        ITTR \cite{zheng2022ittr} & 33.6 & - & - &  45.1  &   -  &    -  &  - & - & 73.4 &  -   &   -  & - &   -    &    -   &   -    & - \\ \hline
        LDA-GAN \cite{ZHAO2022355} &  65.7 & 96.7 & 83.1 &  -  &   -  &    -  &   & 148.9 &   -   &  -   &   -  & 165.7 &   -    &    -   &   -    & - \\ \hline
        MSGAN \cite{mao2019mode} &     -  & 53.32 & 46.23 &     -  &     -   & 78.32 &     -  &     -  &    -   & 169.97 & 111.19 &     -  &   -    &    -   &    -   & - \\ \hline
        MUNIT \cite{huang2018multimodal} & 133.8 & 114.08 & 99.14 & 91.4  & 91.4  & 335.72 & 130.55 & 186.88 & 98.58 & 240.11 & 54.52 & 213.77 & 138.86 & 61.1  & 138   & 224.96 \\ \hline
        NEGCUT \cite{NEGCUT} & 39.7  &     -  &  -     & 48.5  & 48.5  &      - &     -  &     -  &     -  &  -     &     -  &    -   &    -   &    -   &  -     & - \\ \hline
        NICE-GAN \cite{chen2020reusing} & 65.9  & 76.03 & 76.44 &      - &  -     &  -     & 122.27 & 169.79 & 124.11 &   -    &  -     & 192.19 & 112   &      - &     -  & - \\ \hline
        NOT \cite{korotin2023neural}  & 104.3 & 185.5 &     -  & 221.3 &  -     &  -     &      - &      - &     -  &     -  &  -     &  -     & -      &    -   &  -     & - \\ \hline
        StarGAN \cite{choi2018stargan} & 150.3 &   -    &  -     &     -  &     -  &     -  & 211.44 &    -   &    -   &     -  & 140.41 &    -   & 36.68 & 68.3  & 117   &  - \\ \hline
        U-GAT-IT \cite{kim2019u} &     -  &   -    &    -   &  -     &      - &     -  &      - &  -     & 95.63 &     -  &  -     &    -   &  -     & 72.2  & 170   & - \\ \hline
        U-GAT-IT-light \cite{kim2019u} & 113.44 & 88.41 & 80.33 &  -     &  -     &    -   & 137.7 & 134.2 &    -   &     -  &   -    & 179.29 & 123.57 &    -   &    -   & - \\ \hline
        UNIT \cite{liu2017unsupervised} & 131.04 & 112.07 & 95.93 & 91.4  & 91.4  & 239.58 & 136.8 & 161.25 &     -  & 213.65 & 90.32 & 195.24 & 98.12 & 62.1  & 155   & 253.98 \\ \hline
        UNIT-DDPM \cite{sasaki2021unit} &     -  & 109.98 & 113.7 &     -  &     -  & 169.95 &   -    &   -    &   -    & 193.06 &  -     &  -     &   -    &     -  &    -   & 116.23 \\ \hline
        UNSB \cite{kim2023unpaired}  & 35.7  & 73.9  &    -   & 53.2  &     -  & -      &  -     &     -  &    -   & -      &  -     & -      &  -     &   -    & -      & - \\ \hline

	\end{tabular}}
	\hspace*{-1cm}
\end{table}

\begin{table}[]	
	\centering
	\caption{KID values of different methods for Partially Content preserving category of tasks. \textbf{Abbreviations:} S: Selfie, An: Anime, M: Map, A:Aerial, Ap: Apple, O: Orange, L: Label, C: Cityscape, PH: Photo, P:Portrai, V: Vangogh, Sm: Summer, W: Winter, H: Horse and Z: Zebra}
\label{Partially_KID}	
    \hspace*{-2cm}
    \scalebox{0.75}{
	\begin{tabular}{|c|c|c|c|c|c|c|c|c|c|c|c|c|c|}
		\hline
		\multicolumn{1}{|c|}{Method/Task} &  H2Z & Ap2O & O2Ap &  Sm2W &  S2An & An2S & PH2V & V2PH & W2Sm & M2A & L2C & PH2P & P2PH   \\ \hline
                		   
        AGGAN \cite{AGGAN} & 6.93  & 10.36 & 4.54 &  -     & 14.63 & 12.72 & 6.95 & 5.85 &  -     &  -     &   -    & 2.33 & 2.19 \\ \hline
        AttentionGAN \cite{tang2021attentiongan} & 2.03  & 10.03 & 5.41 &  -     & 12.14 &  -     &  -     &   -    &  -     &   -    &  -     &   -    & - \\ \hline
        CUT \cite{CUT}   & 0.541 & -      &      - & 1.207 & 25.5 &     -  & 4.81 &     -  &  -     & 3.301 & 1.611 &     -  & - \\ \hline
        CycleGAN \cite{zhu2017unpaired} & 3.24  & 8.48 & 9.82 & 1.022 & 13.08 & 2.29 & 4.75 & 4.68 & 1.36 & 3.43  & 3.532 & 1.84 & 1.82 \\ \hline
        DiscoGAN \cite{kim2017learning} & 13.68 & 18.34 & 21.56 & -      &     -  &     -  & -      &  -     &     -  &     -  & -      &  -     & - \\ \hline
        DistanceGAN \cite{benaim2017one} & 1.856 & -      &  -     & 2.843 & -      & -      &  -     &  -     &  -     & 5.789 & 4.41  &  -     & - \\ \hline
        DRIT \cite{lee2018diverse}  & 7.4   & 9.65 & 6.5 & 1.27  & 15.08 & 14.85 & 5.43 & 5.62 & 1.69 &  -     &  -     & 5.85 & 4.76 \\ \hline
        DualGAN \cite{yi2017dualgan} & 10.38 & 13.04 & 12.42 &  -     &  -     &  -     &  -     &    -   &   -    &   -    &     -  &     -  & - \\ \hline
        FQ-GAN \cite{FQ-GAN} & 2.93  &     -  & -      & -      & 11.4 & 10.23 & 6.54 & 5.21 &     -  &  -     &      - &   -    & - \\ \hline
        GcGAN \cite{fu2019geometry} & 2.051 & -      & -      & 2.755 & 20.92 & 3.77 &    -   &     -  &    -   & 5.153 & 6.824 &  -     & - \\ \hline
        GDGC \cite{cao2023guided}  & 1.61  & 6.21 & 3.4 &       &    -   &    -   &     -  &  -     &      - &    -   &   -    &  -     & - \\ \hline
        IEGAN \cite{ye2021independent} &    -   & 8.7 & 4.65 & 2.07  &  -     &  -     &     -  &   -    & 1.76 &     -  &   -    &  -     & - \\ \hline
        InvolutionGAN \cite{deng2023involutiongan} &   1.0017   &  &  &   &  1.4274  &  -     &    2.6214  &   -    &  &     -  &   -    &  -     & - \\ \hline
        IrwGAN \cite{xie2021unaligned}  & 1.83  &   -    &  -     &  -     & 2.58 & 2.07 & -      &  -     &    -   &  -     &     -  &     -  & - \\ \hline
        LDA-GAN \cite{ZHAO2022355} & 1.4 &   7.33    & 5.21  &  1.61 & -  & - & -   &  -     &   1.22   &  -     &     -  &     -  & - \\ \hline
        MixerGAN \cite{cazenavette2021mixergan} & 6.81  & 10.23 & 4.5 &    -   &  -     &     -  &  -     &  -     &    -   &  -     &    -   &     -  & - \\ \hline
        MUNIT \cite{huang2018multimodal} & 6.92  & 13.45 & 9 & 5.27  & 13.85 & 3.66 & 4.5 & 9.53 & 4.66 & 12.03 & 6.401 & 4.75 & 3.3 \\ \hline
        NICE-GAN \cite{chen2020reusing} & 2.09  & 11.67 & 8 & 0.67  & 4.51 & 2.45 & 3.53 & 2.79 & 1.22 &    -   &  -     &  -     & - \\ \hline
        NOT \cite{korotin2023neural}   & 5.012 &     -  &     -  & 8.732 & -      &    -   &    -   & -      & -      & 16.59 & 19.76 &     -  & - \\ \hline
        SPA-GAN \cite{emami2020spa} & 2.01  & 3.77 & 2.38 & -      &   -    &  -     &   -    &   -    &   -    & -      &  -     &   -    & - \\ \hline
        UAIT \cite{mejjati2018unsupervised}  & 6.93  & 6.44 & 5.32 &    -   &    -   &  -     &     -  &     -  &     -  &    -   &  -     &     -  & - \\ \hline
        U-GAT-IT \cite{kim2019u} & 7.06  & 10.17 &     -  & -      & 11.61 & 11.52 & 4.28 & 5.61 &   -    &  -     &  -     & 1.79 & 1.69 \\ \hline
        U-GAT-IT-light \cite{kim2019u} & 5.13  &     -  & 4.92 & 1.43  &     -  & -      & 6.03 & 4.91 & 1.82 &     -  &     -  &  -     & - \\ \hline
        UNIT \cite{liu2017unsupervised} & 11.22 & 11.68 & 11.76 & 5.36  & 14.71 & 26.32 & 4.26 & 7.44 & 4.63 &     -  &    -   & 1.2 & 1.42 \\ \hline
        UNSB \cite{kim2023unpaired}  & 0.587 &    -   &     -  & 0.421 &  -     &  -     &     -  &     -  &  -     & 2.013 & 1.191 &      - & - \\ \hline
        UVCGAN \cite{torbunov2023uvcgan}  & - &    -   &   -  & - &  1.35 &  2.33  &     -  &     -  &  -     & - & - &    - & - \\ \hline

	\end{tabular}}
	\hspace*{-1cm}
\end{table}

\subsection{Non Content preserving}
 
In this section, we present the results of I2I tasks which are Non Content preserving. For tasks in this category, image translation has to be at the semantic level by changing content information found in the original images. In Fig. \ref{PCP_NCP_samples_fig}(b), we show qualitative results for Glass $\rightarrow{}$ No-Glass and Male $\rightarrow{}$ Female tasks. Similar to the previous two sections for tasks in Non Content preserving category, we show the results of the I2I models on different metrics in separate Tables. The results of the I2I models for metrics FID and KID, are given in Tables \ref{Non_FID} and \ref{Non_KID}, respectively. For instance, on Cat $\rightarrow{}$ Dog task, DSMAP and CycleGAN have the lowest and highest FID among other methods.

\begin{table}[]	
	\centering
	\caption{FID values of different methods for Non Content preserving category of tasks. \textbf{Abbreviations:} Ce: CelebA-HQ, AF: AFHQ\_I2I, K: Keypoint, L: Label, PH: Photo, G: Glass, NG: No-Glass, Ml: Male, Fl: Female, M: Man, W: Woman, H: Human, D: Dog, C: Cat, C2D\_AF: Cat2Dog\_AFHQ, and Wi: Wild}
\label{Non_FID}

    \hspace*{-2cm}
    \scalebox{0.7}{
	\begin{tabular}{|c|c|c|c|c|c|c|c|c|c|c|c|c|c|c|}
		\hline
		\multicolumn{1}{|c|}{Method/Task}  & D2C\_AF & H2C & Ml2Fl & Ce & C2D\_AF & K2PH & L2PH & C2H & Fl2Ml & G2NG & M2W & W2M & D2C &  Wi2D \\ \hline
        
        ACL-GAN \cite{zhao2020unpaired} & -  &  - & 16.63 & -  & -  & -  & -  &  - & -  & 23.72 &  - & -  & -  & - \\ \hline
        CDC\_IT \cite{zhang2020cross} &  - &  - & -  & 14.3 & -  & 14.4 & 26.4 & -  & -  & -  & -  & -  & -  & - \\ \hline
        COCO-FUNIT \cite{saito2020coco} & 97.08 & 236.9 & 39.19 & -  & -  & -  & -  & -  & -  & -  &  - &  - & -  & - \\ \hline
        CoCosNet \cite{zhang2020cross} &  - & -  & -  & 14.3 & -  & 26.9 & 26.4 &  - & -  & -  &  - & -  &  - & - \\ \hline
        CoCosNet v2 \cite{zhou2021cocosnet} &  - &  - &   & 12.85 &  - & 22.5 & 25.2 & -  &  - & -  &  - & -  &  - & - \\ \hline
        CUT \cite{CUT} & 22.79 &  - & 31.94 &  - & 76.2 & -  & -  &  - & -  & 15.5 &  - &  - & -  & 92.94 \\ \hline
        CycleGAN \cite{zhu2017unpaired} & 119.32 & 44.23 & 21.3 & -  & 85.9 & -  & -  & 57.92 & -  & 48.71 & 161.62 & 146.97 & 119.32 & - \\ \hline
        DCLGAN \cite{han2021dual} & -  & 77.22 & 59.65 &  - & 60.7 &  - & -  & 83 & 54.7 &  - & -  &  - & -  & - \\ \hline
        DMIT \cite{yu2019multi} & -  & 228.28 & 71.23 &  - &  - &  - &  - & 225.49 & 94.98 & -  &  - & -  &  - & - \\ \hline
        DRIT \cite{lee2018diverse} & 94.5 & 166.18 & 65.86 & 52.1 & 79.57 &  - &  - & 69.53 & 98.07 & -  & 141.13 & 187.15 & 94.5 & - \\ \hline
        DRIT++ \cite{lee2018diverse} & -  &  - & 24.61 & -  &  - &  - &  - &  - &  - & 33.06 &  - &  - &  - & - \\ \hline
        DSMAP \cite{chang2020domain} & 19.69 & 160.1 & 57.34 & -  & 13.6 & -  & -  & 103.65 & 64.44 & -  & -  & -  &  - & - \\ \hline
        EGSC-IT \cite{ma2018exemplar} &  - &  - & -  & 29.5 &  - & 29 & 168.3 &  - &  - & -  &  - &  - & -  & - \\ \hline
        EGSDE \cite{zhao2022egsde} &  - & -  & 41.93 &  - & 65.82 & -  &  - & -  & -  & -  &  - & -  &  - & 59.75 \\ \hline
        EM-LAST \cite{EM-LAST} & 27.99 &  - & 47.76 &  - & -  & -  &  - & -  & -  &  - &  - & -  & -  & 72.49 \\ \hline
        GcGAN \cite{fu2019geometry} & 153.83 &  - & -  & -  & 96.6 &  - &  - & -  &  - &  - &  - & -  & -  & - \\ \hline
        IrwGAN \cite{xie2021unaligned} & 53.46 & -  & -  & -  & 61 & -  & -  & -  & -  &  - &  - &  - & -  & - \\ \hline
        LETIT \cite{LETIT} & 30.32 & -  & 48.51 & 12.5 & 45.2 & -  &  - &  - &  - & -  &  - &  - & -  & 83.2 \\ \hline
        MSGAN \cite{mao2019mode} & -  &  - & -  & 33.1 & 20.8 & -  & -  & -  &  - &  -  &  - & -  & -  & - \\ \hline
        MUNIT \cite{huang2018multimodal} & 174.32 & 90.5 & 66.61 & 56.8 & 104.4 & 74 & 129.3 & 89.01 & 48.41 & 28.58 & 150.94 & 163.39 & 53.25 & - \\ \hline
        NICE-GAN \cite{chen2020reusing}  & 48.79 & -  & -  & -  & 48.8 & -  &  - &  - &  - & -  & 139.3 & 145.31 & 48.79 & - \\ \hline
        RABIT \cite{zhan2022bi} & -  &  - &  - & 20.44 &  - & 12.58 & 24.35 &  - & -  & -  & -  & -  & -  & - \\ \hline
        SCS-UIT \cite{liu2021separating} & -  & 68.66 & 57.17 & -  & -  & -  &  - & 72.53 & 65.12 &  - & -  & -  & -  & - \\ \hline
        SDEdit \cite{meng2022sdedit} & -  &  - & 49.43 &  - & 74.17 &  - &  - &  - &  - &  - &  - &  - &  - & 68.51 \\ \hline
        SelectionGAN \cite{SelectionGAN} & -  & -  &  - & 42.41 & -  & 38.31 & 35.1 & -  & -  & -  & -  & -  & -  & - \\ \hline
        SPADE \cite{park2019semantic} &  - &  - &  - & 31.5 &  - & 34.4 & 33.9 &  - &  - &  - &  - &  - &  - & - \\ \hline
        StarGAN v2 \cite{choi2020stargan} & 22.08 & 11.35 & 19.8 & 53.2 & 33.9 & 43.29 & 98.72 & -  & -  & -  & -  &  - &  - & 31 \\ \hline
        TransGaGa \cite{wu2019transgaga} & 23.23 & 21.88 & -  &  - & -  &  - & -  & 32.25 &  - &  - & -  &  - & -  & - \\ \hline
        TraVeLGAN \cite{Travelgan} & 58.91 & 85.28 & 66.6 &  - & -  &  - &  - &  - &  - &  - &  - &  - &  - & - \\ \hline
        U-GAT-IT \cite{kim2019u} & 38.31 & 110.57 & 29.47 & -  & -  & -  &  - &  - &  - & 18.1 & -  & -  &  -  & - \\ \hline
        U-GAT-IT-light \cite{kim2019u} & 80.75 &  - & -  &  - & 64.36 &  - & - &  - &  - &  - & 151.06 & 159.73 & 80.75 & - \\ \hline
        UNIT \cite{liu2017unsupervised} & 59.56 & 35.26 & -  & -  & 63.78 & -  &  - & 98.39 &  - & -  & 147.81 & 179.56 & 59.56 & - \\ \hline
        UVCGAN \cite{torbunov2023uvcgan} & - & - & 9.6 & -  & - & -  &  - & - & 13.9 & 14.4 & - & - & - & - \\ \hline
        
	\end{tabular}}
	\hspace*{-1.5cm}
\end{table}

\begin{table}[]	
	\centering
	\caption{KID values of different methods for Non Content preserving category of tasks. \textbf{Abbreviations:} D: Dog, C:Cat,  Ml: Male, Fl: Female,  M: Man, W: Woman, Ls: Lion\_sealion, Tt: Tigert\_tigerbeetle, C2D\_AF: Cat2Dog\_AFHQ, and D2C\_AF: Dog2Cat\_AFHQ}
\label{Non_KID}	
	\resizebox{\textwidth}{!}{
	\begin{tabular}{|c|c|c|c|c|c|c|c|c|c|}
		\hline
		\multicolumn{1}{|c|}{Method/Task} & D2C\_AF & C2D\_AF & M2W & W2M & C2D & D2C & Ls2Tt & Tt2Ls & Ml2Fl \\ \hline
            
            ACL-GAN \cite{zhao2020unpaired}  &    -   &     -  &     -  &    -   &   -    &     -  &    -   &   -    & 0.015 \\ \hline
            AGGAN \cite{AGGAN} & 9.45  & 9.84  &    -   &   -    &     -  &     -  &     -  &    -   & - \\ \hline
            CycleGAN \cite{zhu2017unpaired} & 4.93  & 6.93  & 56.4  & 48.4  & 69.3  & 49.3  & 4.75 & 6.95 & 0.021 \\ \hline
            DRIT \cite{lee2018diverse} & 5.2   & 10.92 & 35.3  & 100   & 4.57  & 52    &     -  &  -     & -  \\ \hline
            EM-LAST \cite{EM-LAST} & 0.82  & 3.05  &    -   &  -     &      - &     -  &     -  &  -     & 1.75 \\ \hline
            GcGAN \cite{fu2019geometry} & 8.71  & 2.47  &    -   & -      &    -   &    -   & 8.83 & 6.48 & - \\ \hline
            IEGAN \cite{ye2021independent} &   -    &   -    & 39.1  & 35.6  & 6.7   & 9.5   &   -    &      - & - \\ \hline
            IrwGAN \cite{xie2021unaligned} & 1.84  & 2.07  &  -     & -      &    -   &   -    & 2.44 & 2.34 & - \\ \hline
            LETIT \cite{LETIT} & 1.01  & 3     &   -    &    -   &    -   &    -   &      - &   -    & 1.9 \\ \hline
            MUNIT \cite{huang2018multimodal} & 1.26  & 5.77  & 56.2  & 66.8  & 24.2  & 12.6  & 26.33 & 25.35 & 0.019 \\ \hline
            NICE-GAN \cite{chen2020reusing} & 1.58  & 1.2   & 31.2  & 42.8  & 12    & 15.8  & 6.98 & 11.99 & - \\ \hline
            U-GAT-IT \cite{kim2019u} & 8.15  & 7.07  &    -   &  -     &      - &   -    &   -    &  -     & - \\ \hline
            U-GAT-IT-light \cite{kim2019u} & 3.22  & 2.49  & 43.1  & 57.5  & 24.9  & 32.2  &   -    &  -     & - \\ \hline
            UNIT \cite{liu2017unsupervised}  & 1.94  & 1.94  & 48.1  & 91.8  & 19.4  & 19.4  &   -    &   -    & - \\ \hline

	\end{tabular}}
\end{table}

	\section{Sim2Real Benchmark}
	Transferring simulation to real images can be considered as an application to fully content preserving or partially content preserving unsupervised image to image translation methods. While producing large realistic data-sets can be challenging, this problem can effectively ensure that training models learn from data that resembles the environment they will encounter in real applications. Some application to approaches contributing simulation to real image translation can be denoted as Medical Imaging, Autonomous driving, robotics, detection task with low supervised data and making art. 
	
	Simulation to real problem can be beneficial to machine learning task in several aspects. One contribution of simulation to real image translation  to the machine learning real-world scenarios is data augmentation in applications in which large and diverse data-set preparation can be challenging or costly. In this case, providing more varied data for training in an specific domain can improve robustness and generalization power and decrease the chance of over-fitting in machine learning models. Another contribution can be increasing training performance by providing more realistic simulated data for real-world application. In addition, reducing risk of errors associated with real-world experiments enhance safety and reliability of the model predictions. further, deploying a simulation to real image translation approach before exposing the model to the real-world applications helps researchers to experiment new  algorithms in a more restricted, controlled environment. 
	
	In this section, we will explain available data-sets and generated data-sets for simulation to real problem alongside evaluations done upon some of the selected papers for implementation in the previous sections in this paper.
	
	\subsection{Benchmark data-set for simulation to real benchmark}
	In this section, A benchmark data-set including source and target data-sets specifically made for content-preserving simulation to real methods is introduced.
	
	For utilizing datasets in real  domain of translation, we combined Visdrone, MIO and VeRi datasets with images labeled as Pickup, Car, Bus, Van and truck classes. Also we used a data-set of simulated images of Pickup, Car, Bus, Van and truck as the second domain.
	
\begin{figure}[!tp]
		\centering
		\includegraphics[width=0.6\textwidth]{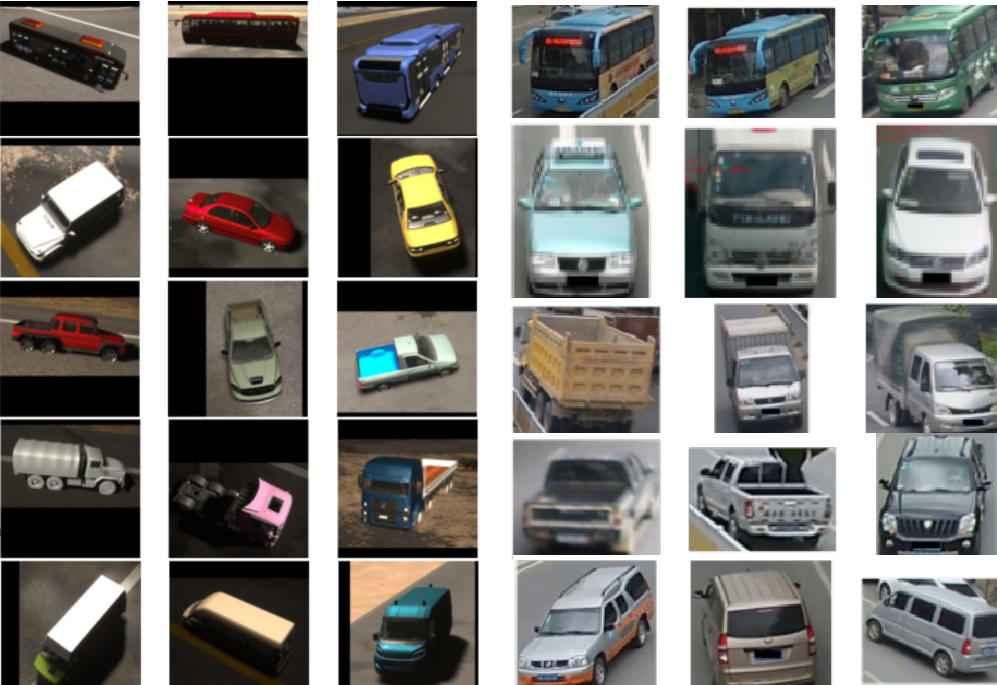}
		\caption[]{Simulation sample images (first row) and Real sample images (second row). Each line of images corresponds to 5 classes Bus, car, pickup, truck and van respectively.}
		\label{fig:simulation-real-datasets}
	\end{figure}

    For the real domain, each data set is split into high-resolution and low-resolution sets. The threshold of 65535 pixels or images in size (256, 256) is used to split the dataset. All images with pixel sizes lower than the threshold are then resized to (128, 128) and padded randomly if necessary. Images with pixel size higher than the threshold are then resized to (256, 256). In Table \ref{tab:low_high_real}, brief information on the number of images in the real domain in each class is reported. Images from 5 classes of simulation data set have been randomly sampled and demonstrated in Figure \ref{fig:simulation-real-datasets}.
	
    For the simulation domain, we regularized the image sizes of the simulation data set with the real data set. Low resolution subset contains images with (128, 128) sizes, and high-resolution data set contains images with (256, 256) sizes.  In Table \ref{tab:low_high_sim}, brief information on the number of images in the simulation domain in each class is reported. In Figure \ref{fig:simulation-real-datasets}, images of 5 classes in the VeRi data set have been randomly sampled and demonstrated.
	
	\begin{table}[h]
	    \caption{Image distribution and resolution for each class for datasets MIO, Visdrone and Veri}
		\label{tab:low_high_real}
		\centering
		\resizebox{\textwidth}{!}{
		{}{%
			\begin{tabular}{|l|l|l|l|l|l|l|l|l|l|l|}
				\hline
				& \multicolumn{2}{c|}{\textbf{Pickup}} & \multicolumn{2}{c|}{\textbf{Car}} & \multicolumn{2}{c|}{\textbf{Bus}} & \multicolumn{2}{c|}{\textbf{Van}} & \multicolumn{2}{c|}{\textbf{Truck}} \\ \hline
				\textbf{} & Low & High & Low & High & Low & High & Low & High & Low & High \\ \hline
				\textbf{MIO} & 48253&2653&255954&4564&6026&4290&8956&723&4067&1053 \\ \hline
				\textbf{Visdrone} &328&3&53776&556&4207&147&20924&328&6878&315\\ 
				\hline
				\textbf{VeRi} & -&3094&-&35525&-&3134&-&2323&-&5249 \\ \hline
			\end{tabular}%
		}}
	\end{table}
	
	\begin{table}[h]
        \tiny
		\caption{Simulation Data distribution and resolution for each class}
		\label{tab:low_high_sim}
		\centering
		\resizebox{\textwidth}{!}{%
			\begin{tabular}{|l|l|l|l|l|l|l|l|l|l|}
				\hline
				\multicolumn{2}{|c|}{\textbf{Pickup}} & \multicolumn{2}{|c|}{\textbf{Car}} & \multicolumn{2}{|c|}{\textbf{Bus}} & \multicolumn{2}{|c|}{\textbf{Van}} & \multicolumn{2}{|c|}{\textbf{Truck}} \\ \hline
				Low & High & Low     & High & Low & High & Low & High & Low & High \\ \hline
				37 &  4774 & 1245 &10275 &- &2392&362&2491&10&4548  \\ \hline
			\end{tabular}%
			}
	\end{table}

	\subsection{Evaluation Results}
    As said already, we have created a simulation to real benchmark (Sim2Real) in which we assess content-preserving approaches on our simulation to real benchmark data set. In the following sections, we dive into the results of selected papers on various metrics introduced in section \ref{Metrics_sec} and we present different configuration setups alongside visual and metric performance comparisons. We have selected 8 models, CycleGAN \cite{zhu2017unpaired}, DRIT \cite{lee2018diverse}, GcGAN \cite{fu2019geometry}, StyleFlow \cite{fan2022styleflow}, SRUNIT \cite{jia2021semantically}, VSAIT \cite{theiss2022unpaired} an UNSB \cite{kim2023unpaired} from previous sections in this paper that seemed to have parameters to be set for preserving image content as in their transformation from domain A and domain B, and then evaluated these models on our benchmark dataset.  
	
	\subsubsection{CycleGAN}
\begin{figure}[!h]
		\centering
		\includegraphics[width=0.5\textwidth]{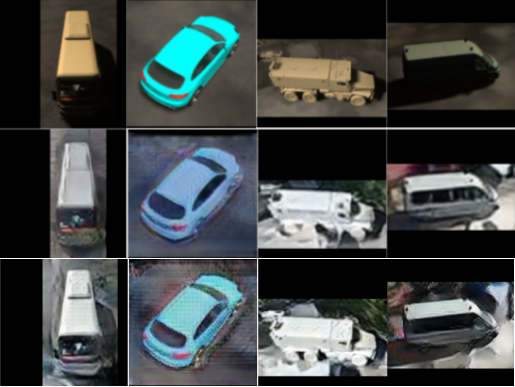}
		\caption[]{Visual results of experimenting CycleGAN on Sim2Real benckmark dataset. First row is a set of sample simulation images (a column for each class). The second and third row is transferred images from simulation (first row) to real with parameters batch-size=4, $\lambda_A=10$, $\lambda_B=10$ and batch-size=32, $\lambda_A=10$, $\lambda_B=10$ respectively.}
		\label{fig:cyclegan}
	\end{figure}

    We evaluated CycleGAN \cite{zhu2017unpaired}, introduced in previous sections, on batch size, $\lambda_A$ and $\lambda_B$, the most effective hyperparameters controlling content preservation, and also various evaluation metrics including FID, IS, NDB, JSD and LPIPS. In the following, we discuss different test and the results for each combination of aforementioned hyperparameters.
    We trained and tested CycleGAN on Sim2Real dataset with the input image sizes of (128, 128) for 20 epochs and batch size of 4. Then for experimenting on batch size effect on the results, we trained and tested the model for batch size 32. In both of the experiments $\lambda_A$ and $\lambda_B$ were set to 10. We have also experimented  the effect of hyperparametrs $\lambda_A$ and $\lambda_B$ on the model performance and set $\lambda_A$ and $\lambda_B$ to 5 to see the visual results and evaluation metrics. In table \ref{tab:sim2real_metrics_papers} result of testing model on Sim2Real benchmarck on FID, IS, NDB, JSD and LPIPS metrics are reported. From this table we can conclude changing hyperparameter $\lambda$ from 10 to 5 do not increase the performance results. In addition, visual reports of different experiments are shown on figure \ref{fig:cyclegan}. By checking more translated images of these tests we can conclude visual results of experiments with batch size 32 have more quality than experiment with batch size 4. Furthermore, visual results degrade when using $\lambda$ 5 instead of 10. 
    \begin{table}[!ht]
    	\centering
        \caption{CycleGAN Results for benchmark sim2real data-set on metircs}
        \label{tab:sim2real_metrics_papers}
        \resizebox{\textwidth}{!}{
        \small
        \begin{tabular}{|c|c|c|c|c|c|}
            \hline
            Run\textbackslash{}Metric & FID $\downarrow$ & IS $\uparrow$ & NDB $\downarrow$ & JSD $\downarrow$ & LPIPS $\downarrow$ \\ \hline
            Reference metrics        & 71.36    & 5.57   & 27  & 0.15   & NA      \\ \hline
            CycleGAN, b 4, $\lambda_A=10$, $\lambda_B=10$  & 34.15    & 4.702  & 8   & 0.0429 & 0.6181 \\ \hline
            CycleGAN, b 32, $\lambda_A=10$, $\lambda_B=10$ & 40.42    & 4.8605 & 12  & 0.0532 & 0.6161 \\ \hline
            CycleGAN, b 32, $\lambda_A=5$, $\lambda_B=5$  & 110.4867 & 4.092  & 6   & 0.0406 & 0.6135 \\ \hline
            CycleGAN, b 32, $\lambda_A=20$, $\lambda_B=20$ & 56.2949  & 4.7894 & 15  & 0.0595 & 0.6115 \\ \hline
            StyleFlow, B1   & 128.76 &3.39 &44 &0.23 & -\\ \hline
            StyleFlow, B4   &59.34 &4.69&26& 0.1 & -\\ \hline
            StyleFlow, B8    & 62.59 &4.5& 20& 0.1 & -\\ \hline
            StyleFlow, B16   & 58.87& 4.84& 25 &0.11 & -\\ \hline
            StyleFlow, P1     & 76.23& 4.69 &33 &0.12 & -\\ \hline
            StyleFlow, P2     &76.25 &4.69& 26& 0.11 & -\\ \hline
            StyleFlow, P3     &85.68 &4.04 &36 &0.16 & -\\ \hline
            StyleFlow, P4     &84.11 &5.1 &27 &0.14 & - \\ \hline
            GcGAN, b 4, $\lambda_A=10$, $\lambda_B=10$, $\lambda_i=0.7$   & 54.3608    & 5.1081  & 25   & 0.22 & - \\ \hline
            GcGAN, b 12, $\lambda_A=10$, $\lambda_B=10$, $\lambda_i=0.5$ & 48.3213    & 5.6685 & 17  & 0.18 & - \\ \hline
            SRUNIT, Test 1                & 66.5090 & 5.4351 & 19  & 0.809 & 0.3183 \\ \hline
            SRUNIT, Test 2                & 46.8759 & 5.1309 & 18  & 0.701 & 0.3734 \\ \hline
            SRUNIT, Test 3                & 57.3073 & 5.1994 & 15  & 0.792 & 0.2864 \\ \hline
            SRUNIT, Test 4                & 57.3073 & 4.9113 & 20  & 0.1261 & 0.1952 \\ \hline
            VSAIT, b 4                & 60.05 & 5.08 & 29  & 0.134 & 0.053 \\ \hline
            VSAIT, b 32               & 51.26 & 5.22 & 19  & 0.1   & 0.091 \\ \hline
            UNSB, ${\lambda_s}_b=0.1$, Input image size=(128, 128) & 64.9163 & 4.6683 & 18 & 0.879 & 0.3664 \\ \hline
            UNSB, ${\lambda_s}_b=1$, Input image size=(256, 256) & 57.2819 & 4.2189 & 26 & 0.1009 & 0.3294 \\ \hline
        \end{tabular}}
    \end{table}
    
    \subsubsection{StyleFlow}
\begin{figure}[!h]
	\centering
	\small
	\includegraphics[width=0.6\textwidth]{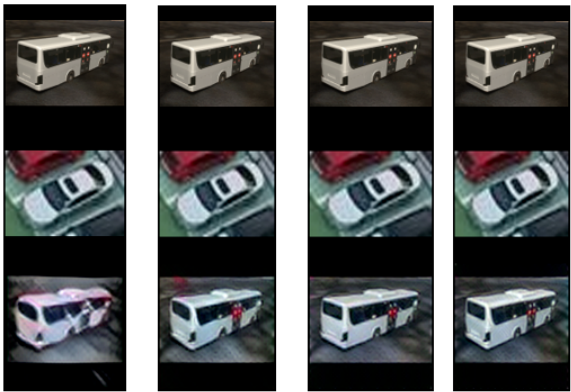}
	\caption[]{StyleFlow visual results for hyperparameter combination B1, B4, B8 and B16 (respectively demonstrated in columns from left to right). First row is the simulation image, second row is the randomly selected style image from real images and the third row is the translated real image generated by StyleFlow.}
	\label{fig:st_batch}
    \end{figure}

\begin{figure}[!h]
	\centering
	\includegraphics[width=0.6\textwidth]{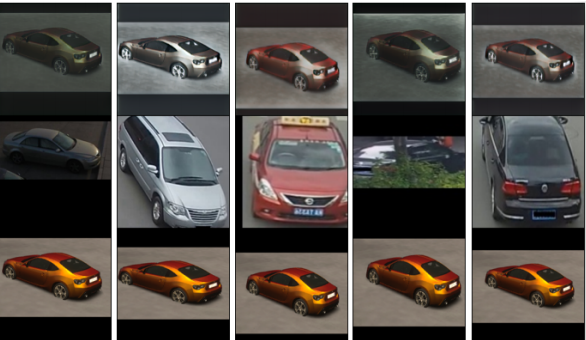}
	\caption[]{StyleFlow visual results for a fixed configuration and different input style images. Third row is the simulation image, second row is the randomly selected style image from real images and the first row is the translated real image generated by StyleFlow.}
	\label{fig:st_style}
    \end{figure}

\begin{figure}[!h]
	\centering
	\includegraphics[width=0.5\textwidth]{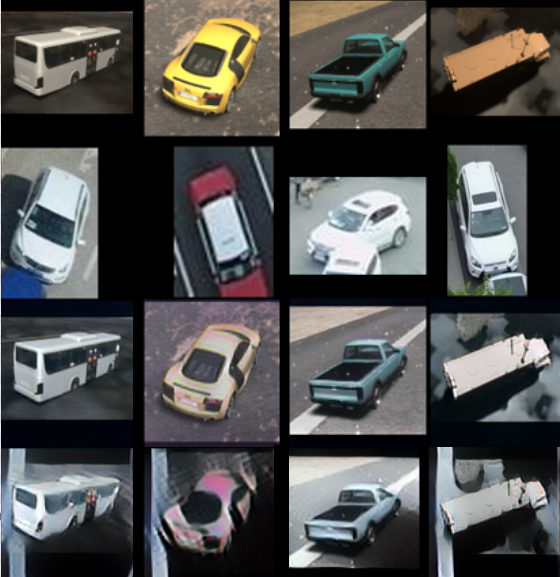}
	\caption[]{StyleFlow visual results for hyperparameter combination P1 and P3 explained in Table \ref{tab:styleflow_tuning_param}.  First row is the simulation image, second row is the randomly selected style image from real images and the third and forth row is the translated real image generated by StyleFlow for P1 and P3 configurations.}
	\label{fig:styleflow}
    \end{figure}

    We tested the StyleFlow \cite{fan2022styleflow} on the Sim2Real benchmark. Selecting a real image for extracting style features plays an important role in the training phase of this model. For selecting the best hyperparameters of this model, we evaluated this model on hyperparameters explained in table \ref{tab:styleflow_param}, and for each combination of hyperparameters shown in table \ref{tab:styleflow_tuning_param}, we reported the evaluation metric results in table \ref{tab:sim2real_metrics_papers}. In table \ref{tab:styleflow_tuning_param}, columns B1, b4, B8, and  B16 demonstrate the batch sizes 1, 4 and 8. For columns P1, P2, P3 and P4,  hyperparameters style\_weight, content\_weight and keep\_ratio are changed as reported. For visual experiments, we choose a single simulation image and a fixed style image and demonstrate translated images with the StyleFlow model using different batch sizes to see the effect of batch size in the output of the model in figure \ref{fig:st_batch}. It can be seen B16 configuration is slightly better than other configurations as it has minimum deformation. Also, in figure \ref{fig:st_style}, we aim to experiment with the effect of different styles in a fixed configuration for StyleFlow. Finally, we have demonstrated images generated in the real domain by this model for P1, P3 in figure \ref{fig:styleflow} to have an overall understanding of result for each configuration introduced in Table \ref{tab:styleflow_tuning_param}.
    
    
    \begin{table}[htbp]
	\centering
	\caption{StyleFlow hyperparameters.}
	\resizebox{\textwidth}{!}{
	\begin{tabular}{|c|c|c|}
        \hline
		\textbf{Parameters} & \textbf{Default value} & \textbf{Details} \\
		\hline
		lr & $0.0001$ & learning rate \\
		\hline
		style\_weight & $1$     & weighting factor employed to balance content and style \\
		\hline
		content\_weight & $0.1$   & weighting factor employed to balance content and style  \\
		\hline
		keep\_ratio & $0.6$   &  weighting parameter for Aligned-Style Loss in the paper  \\
		\hline
		batch size & $1$     & batch\_size \\ \hline
	\end{tabular}
	}%
	\label{tab:styleflow_param}%
\end{table}%
    
    \begin{table}[htbp]
	\centering
	\caption{Hyperparameter selection for each configuration name B1, B2, b4, B8, B16, P1, P2, P3 and P4.}
	\begin{tabular}{|c|c|c|c|c|c|c|c|c|}
		\hline
		\textbf{Hyperparameters} & \textbf{P4}& \textbf{P3} & \textbf{P2} & \textbf{P1} & \textbf{B16} & \textbf{B8}& \textbf{B4} & \textbf{B1} \\
		\hline
		 style\_weight  & 5 & 1  & 1    & 0.5  & 1   & 1   & 1    &1 \\
		\hline
		content\_weight  & 0.1 & 1    & 0.5 & 0.5  & 0.1  & 0.1  & 0.1 & 0.1  \\
		\hline
		keep\_ratio   & 0.6  & 1  &0.6  & 0.5  & 0.6  & 0.6 & 0.6  & 0.6 \\
		\hline
		batch\_size & 1   & 1   & 1 & 1  & 16  & 8  & 4  & 1\\
		\hline
	\end{tabular}%
	\label{tab:styleflow_tuning_param}%
\end{table}%
    \subsubsection{GcGAN}
\begin{figure}[]
		\centering
		\includegraphics[width=0.5\textwidth]{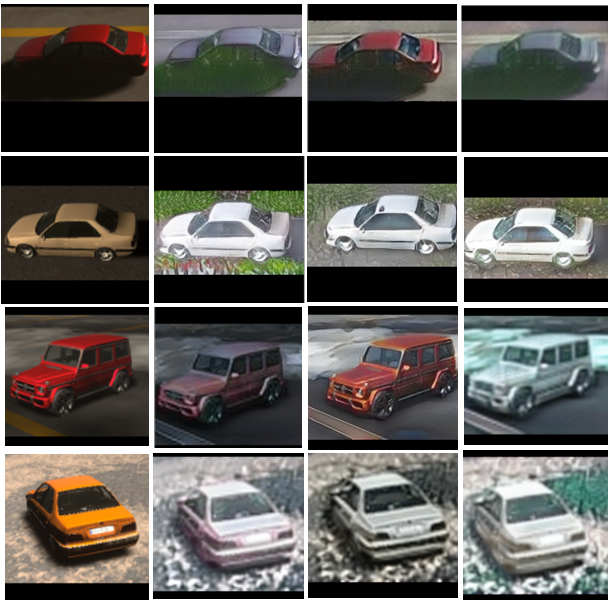}
		\caption[]{Effect of different values for GcGAN hyperparameters $\lambda_A$ and $\lambda_B$ on translating simulation  images to real images. Frist column from left is simulation image and remaining columns from left to right are columns demonstrating translation results for ($\lambda_A=5$ and $\lambda_B=5$), ($\lambda_A=10$ and $\lambda_B=10$) and ($\lambda_A=20$ and $\lambda_B=20$) respectively.}
		\label{fig:gcgan_lambda}
	    \end{figure}

        GcGAN \cite{fu2019geometry} is also one of the previously introduced content-preserving generative adversarial models tested on the Sim2real benchmark. We have evaluated this model on the most important hyperparameters mentioned in the paper. Batch size, $\lambda_A$, $\lambda_B$, and $\lambda_i$ (identity loss coefficient) were the most effective hyperparameters controlling content preservation ability and various evaluation metrics like FID, LPIPS, etc. In the following, we discuss the different tests and the results for each combination of the aforementioned hyperparameters.
        Similar to other tests, we trained and tested CycleGAN on the Sim2Real dataset with the input image sizes of (128, 128). For the first experiment, we aimed to evaluate the effect of different values for hyperparameters $\lambda_A$, $\lambda_B$, and $\lambda_i$. We first calculated FID and IS evaluation metrics for different values of parameters $\lambda_A$, $\lambda_B$, then selected the best value which was 10. In the next step, we experimented with the effect of parameter $\lambda_i$ for each batch size, and the best value of $\lambda_i$ for each batch size was selected (0.7 for batch size 4 and 0.5 for batch size 12). In table \ref{tab:sim2real_metrics_papers}, the best values for evaluation metrics for each batch size are reported. As it can be inferred from the reported results, increasing batch size from 4 to 12 does not increase the performance results. In figure \ref{fig:gcgan_lambda} we have shown the effect of different values 5, 10, and 20 for hyperparameters $\lambda_A$ and $\lambda_B$. We can see that the best value for $\lambda_A$ and $\lambda_B$ seems to be 10 since these values produce more realistic images alongside preserving most of the content of the input images.
    \subsubsection{VSAIT}
\begin{figure}[!h]
	\centering
	\includegraphics[width=0.6\textwidth]{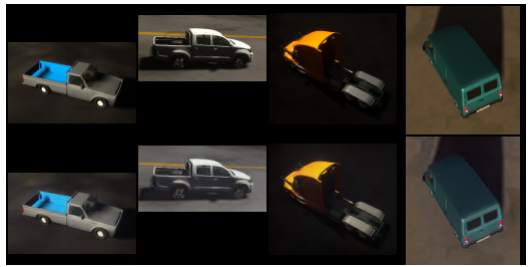}
	\caption[]{Visual results of experimenting VSAIT model on Sim2Real benckmark dataset for batch size 32}
	\label{fig:vsait_b32}
    \end{figure}

       VSAIT \cite{theiss2022unpaired} is the next content-preserving generative adversarial model tested on the Sim2real benchmark. We trained and tested VSAIT on the Sim2Real dataset with the input image sizes of (128, 128) for 20 epochs for all experiments. For this model, we considered batch size hyperparameters to compare model performance on the Sim2Real benchmark dataset. In table \ref{tab:sim2real_metrics_papers}, we have reported evaluation metrics FID, IS, NDB, JSD, and LIPS on the Sim2Real benchmark. Regarding FID, NDB, and JSD, the VSAIT model trained with batch size 32 is the best model. Also, in figure \ref{fig:vsait_b32} VSAIT results for batch size 32 is demonstrated. The first row of each figure is simulation images and the second row is the real images generated with this model.
\subsubsection{SRUNIT}
\begin{figure}[!h]
	\centering
	\includegraphics[width=0.6\textwidth]{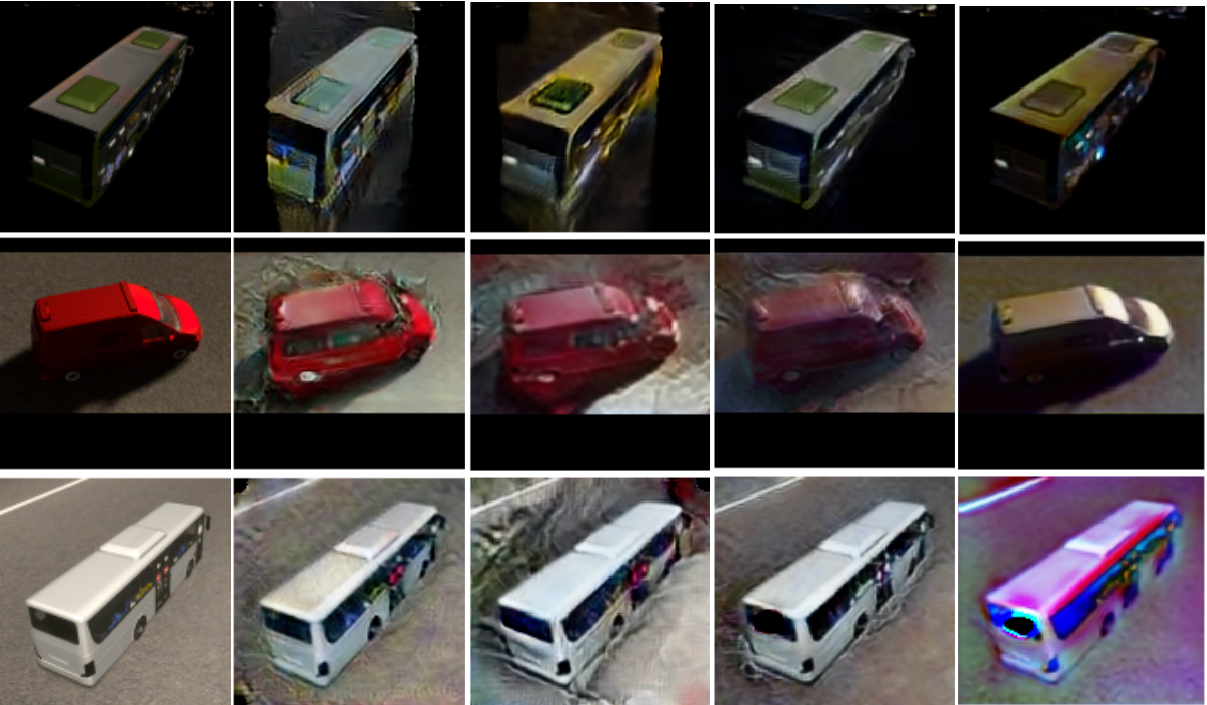}
	\caption[]{Visual results of experimenting with the SRUNIT model on the Sim2Real benchmark dataset for different hyperparameter settings. From left to right, columns are input simulation images and translated images for Test 1, Test 2, Test 3, and Test 4 respectively. Test configuration details are presented in the Table \ref{tab:srunit_param} }
	\label{fig:srunit}
    \end{figure}

    SRUNIT \cite{jia2021semantically} is one of the GAN-based models tested on our benchmark. Like other models, we trained and tested VSAIT on the Sim2Real dataset with the input image sizes of (128, 128) for 20 epochs for all experiments. Visual results of the SRUNIT model can be seen in figure \ref{fig:srunit}. For the default settings, as we can see from this figure, translated images are corrupted and it seems the models have performed weakly in preserving content and image quality. So, we tried to run the model for different subsets of hyperparameters shown in table \ref{tab:srunit_param}. The evaluation metrics of different configurations for this model are reported in Table \ref{tab:sim2real_metrics_papers}.
    
    \begin{table}[htbp]
	\centering
	\caption{SRUNIT hyperparameters.}
	\resizebox{\textwidth}{!}{
	\begin{tabular}{|c|c|c|c|c|c|}
        \hline
		\textbf{Hyperparameter} & \textbf{Definition} & \textbf{Test 1} & \textbf{Test 2} & \textbf{Test 3}  &\textbf{Test 4} \\
		\hline
		${{lambda_G}_A}_N$ &  Adversarial loss coefficient & 5&7&1&1 \\
		\hline

		${{lambda_G}_A}_N$ & Contrasive loss coefficient &10&10&10&10 \\
		\hline
		reg &  Regularization loss coefficient  &  10\% & 10\% & 10\% & 50\%  \\
		\hline
		\end{tabular}
		}%
	\label{tab:srunit_param}%
\end{table}%

\subsubsection{UNSB}
\begin{figure}[!h]
	\centering
	\includegraphics[width=0.6\textwidth]{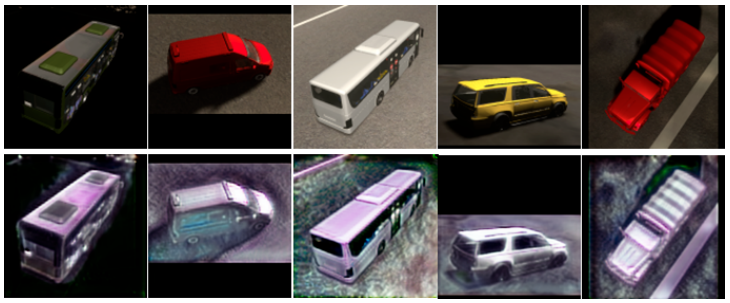}
	\caption[]{UNSB model results.}
	\label{fig:unsb}
    \end{figure}

    For evaluating the UNSB \cite{kim2023unpaired} model on our Sim2Real benchmark, we considered default settings and reported the results on the Sim2Real dataset. Due to the low performance on this dataset, we changed the default settings and changed the input image size to (256, 256). Also changed ${\lambda_s}_b$ in the loss function of the model from 0.1 to 1 and reported the results in table \ref{tab:sim2real_metrics_papers}. Also in figure \ref{fig:unsb}, the visual results for the second test configured in this paper are reported.
\section{Conclusion}

This paper provides a comprehensive review of image-to-image translation algorithms, which we categorized them based on their architectures. Besides, we provided another categorization for GAN-based models based on the characteristics of them. An important challenge in the field of image-to-image translation is to preserve the content of the input image, and the methods presented in some related articles are proposed with the same goal. In this paper, we focus on this issue, image content preservation. In addition, existing and well-known tasks and datasets, evaluation metrics and evaluation results of different methods are summarized in three categories: Fully Content preserving, Partially Content preserving, and Non-Content preserving. This review paper introduces a Sim2Real benchmark and results of some state-of-the-art I2I methods for this defined task.

\end{document}